\documentclass[sigconf]{acmart}

\newif\ifhighlightchanges
\highlightchangesfalse

\ifhighlightchanges
  \newcommand{\rev}[1]{\textcolor{magenta}{#1}}
\else
  \newcommand{\rev}[1]{#1}
\fi

\newcommand{\name}{\textit{StitchOver}}
\newcommand{\stitch}{\textit{JumpStitch}}
\newcommand{\stitches}{\textit{\stitch{}es}}

\newcommand{\shortstitches}{\textit{ShortStitch}es}

\newcommand{\underlaystitches}{\textit{UnderlayStitch}es}

\newcommand{\jointstitches}{\textit{JointStitch}es}

\AtBeginDocument{%
  }

\setcopyright{acmlicensed}
\copyrightyear{2018}
\acmYear{2018}
\acmDOI{XXXXXXX.XXXXXXX}

\acmConference[Conference acronym 'XX]{Make sure to enter the correct
  conference title from your rights confirmation emai}{June 03--05,
  2018}{Woodstock, NY}
\acmISBN{978-1-4503-XXXX-X/18/06}

\copyrightyear{2026}
\acmYear{2026}
\setcopyright{cc}
\setcctype{by}
\acmConference[UIST '26]{The 39th Annual ACM Symposium on User Interface Software and Technology}{November 02--05, 2026}{Detroit, MI, USA}
\acmBooktitle{The 39th Annual ACM Symposium on User Interface Software and Technology (UIST '26), November 02--05, 2026, Detroit, MI, USA}
\acmDOI{10.1145/3830398.3830717}
\acmISBN{979-8-4007-2856-3/2026/11}

\begin{document}

\title{StitchOver: Technical Embroidery on Seamed Fabrics}


\author{Zekun Chang}
\email{zc247@cornell.edu}
\affiliation{%
  \institution{Cornell Tech}
  \city{New York}
  \country{USA}}

\author{Tianhong Catherine Yu}
\email{ty274@cornell.edu}
\affiliation{%
  \institution{Cornell University}
  \city{Ithaca}
  \country{USA}}

\author{Yixuan Gao}
\email{yg478@cornell.edu}
\affiliation{%
  \institution{Cornell Tech}
  \city{New York}
  \country{USA}}

\author{Thijs Roumen}
\email{thijs.roumen@cornell.edu}
\affiliation{%
  \institution{Cornell Tech}
  \city{New York}
  \country{USA}}

\settopmatter{authorsperrow=4}
\renewcommand{\shortauthors}{Chang et al.}

\begin{abstract}
Smart textiles embed interactivity into everyday garments, supporting use cases like always-available sensing for medical applications or sports. Machine embroidery allows integrating functionalities into existing textiles. However, embroidering onto real-world textile goods remains challenging. Textile goods are rarely made of a single homogeneous substrate of fabric, and embroidery with functional materials such as conductive threads requires machines to be more tightly calibrated than for decorative embroidery. In particular, seams, which bring together different substrates, along with machine variability, cause shifts in tension and friction between the functional thread and the textile substrate that frequently lead to defects (70\% of samples in our evaluation).

We present a technique to reliably embroider on seamed fabric even when using functional threads. Our software tool automatically digitizes user-defined stitch patterns by introducing what we call “\stitches{}” to bypass seam interference.

We evaluated our approach under varying machine states (under-tensioned, well-calibrated, and over-tensioned), and across multiple seam and pattern configurations. Our results show that the \stitch{} mechanism eliminates defects, while maintaining conductivity compared to 70\% defects without \stitches{}, and even in poorly calibrated machine states continues to work well. 

\end{abstract}



\begin{teaserfigure}
  \includegraphics[width=\textwidth]{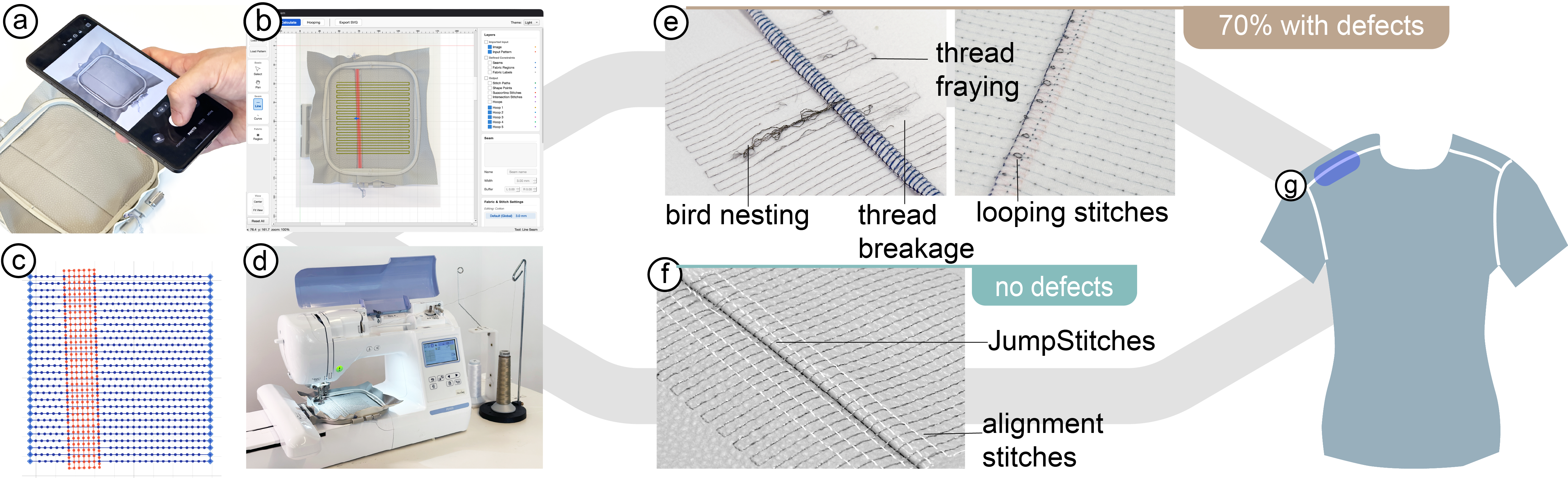}
  \caption{ \rev{Overview of \name{}. (a--d) The workflow: (a) scanning a hooped fabric sample to capture its seam geometry, (b) annotating the seam, (c) generating seam-aware stitch paths, and (d) fabricating the pattern using an embroidery machine. (e) Representative defects that can occur when embroidering across seams using traditional digitization, including bird nesting, thread fraying, thread breakage, and looping stitches. (f) Seam-crossing embroidery produced using \name{}, showing continuous stitch paths without these defects. (g) An example of a garment application requiring embroidered sensors to span a shoulder seam.}}
  \Description{teaser figure}
  \label{fig:teaser}
\end{teaserfigure}



\maketitle

\section{Introduction}

Smart textiles have the potential to transform everyday textiles into platforms for sensing, actuation, and interaction, supporting applications such as healthcare, performance tracking, robotics, and entertainment~\cite{sanchez2021textile}. Common textile fabrication methods such as weaving~\cite{sun2020weaving} and knitting~\cite{zhang2025magtex} create new textiles from scratch. Machine embroidery allows augmenting existing textiles by adding functionality to them. It enables precise position control over stitch placement while providing high freedom in thread routing. This allows threads to be laid out along predefined paths on a variety of textile substrates. Leveraging this programmability, machine embroidery has been widely adopted for making smart textiles~\cite{luo2024adaptive, jiang2024embrogami, nabil2019seamless, preindl2020sonoflex, nabil2021soft}.
It enables the integration of functional threads such as conductive yarns directly into fabrics, supporting the fabrication of textile-based electronic components, including sensing electrodes~\cite{shinoda2025equivalent}, conductive traces~\cite{hamdanSketchStitchInteractiveEmbroidery2018}, antennas~\cite{liu2019embroideryRFID}, and shape-changing~\cite{chang_oristitch_2025}. However, applying technical embroidery to existing textile goods remains challenging. Textile goods are rarely made of a single substrate. They are typically constructed through multi-panel assembly, in some cases involving multiple materials. These panels are joined through seams, resulting in a structure that is inherently non-uniform. Consequently, real textile goods exhibit structural variation across their surface. 

Even on flat, homogeneous fabrics, functional threads, such as silver-coated threads, exhibit higher surface friction and heightened sensitivity to tension variation~\cite{yu2023skinergy}, making the embroidery process more prone to failure compared to conventional threads. 

\rev{This challenge is further amplified with heterogeneous structural variations in seamed soft goods.}
Seams introduce local variations in the material, which in turn create tension and friction changes among the conductive thread, bobbin threads, and the fabric substrates. This mechanical variance leads to failures in embroidery such as thread fraying, bird nesting, needle breakage, and in worse cases, machine damage. \rev{However, the embroidery process cannot simply avoid seams because doing so significantly limits the available embroidery area for functional integration.} While machine maintenance can temporarily mitigate such failures by setting a well-calibrated state, this condition is inherently transient and frequently taking apart the machine to adjust a feature is impractical and uncommon.

For other materials, digital fabrication researchers have characterized similar material variances and provided robust solutions that are independent of the machine state~\cite{roumen_kerf-canceling_2020, roumen_portable_2023}. In this work, we apply such reasoning to technical embroidery, making it work reliably across seams
\rev{(assuming that difficult-to-hoop regions are handled using established embroidery practices).}

We propose inserting what we call “\stitches{}”. This mechanism redesigns needle point placement so that conductive threads avoid penetrating seams, while additional supporting stitches fix the thread on the fabric. Building on this seam-aware mechanism, we implement a digitization tool (\name{}) that takes a circuit embroidery pattern and a hooped textile layout as input. The tool analyzes the spatial relationship between conductive routing and seams, automatically adjusts needle penetration points, and generates supporting stitches. 

Our technical evaluation characterizes the impact of seams on technical embroidery. We find that with \stitches{}, defects are eliminated, compared to 7/10 defective samples without \stitches{}. And even under variations in tension, \stitches{} demonstrate significant improvement over conventional embroidery in introduced defects and conductivity of embroidered circuits. We see this work as a step towards reliable technical embroidery on everyday textiles, a crucial advance for this technology to really reach the broad public.
\section{Background}

\paragraph{Technical Embroidery.} Machine embroidery, traditionally used for textile decoration, is increasingly used as a fabrication technique to create \textit{functional} textiles \cite{Mecnika_Hoerr_Krievins_Jockenhoevel_Gries_2015, Topher_Anderson_2020}. In this paper, we refer to technical embroidery as a process to integrate sensing or actuation in existing textiles using a computerized embroidery machine. However, technical embroidery remains a challenging fabrication process, where failures such as thread breakage or stitching errors can easily occur, leading to failures of the intended functionality~\cite{luo_embroiderer_2023, hamdanSketchStitchInteractiveEmbroidery2018, aigner_embroidered_2020, pointnerEmbroideringResonantCircuits2025}.

\paragraph{Thread Continuity in Technical Embroidery.} Unlike decorative embroidery, where broken threads can be reconnected with minimal impact on the visual outcome, many technical embroidery applications rely on continuous thread paths to maintain their functionality. Thread breakage directly leads to functional failure. For example, heat-shrinkable threads for shape transformation~\cite{chang_oristitch_2025}, conductive threads for electrical circuits~\cite{preindl2020sonoflex}, tubes for liquid metals~\cite{Lin_Kim_Achavananthadith_Xiong_Lee_Kong_Ho_2022}, and reinforcement structures~\cite{Poniecka_Barburski_Urbaniak_2022} all require uninterrupted thread structures. Furthermore, these types of threads tend to require more careful handling (e.g., conductive coating may come off a thread when frayed). Consequently, consistency in technical embroidery is a crucial requirement for successful functionality.

\paragraph{Seams in Textile Construction.} \rev{Real-world textile goods rarely consist of completely planar and homogeneous surfaces. Their construction commonly introduces non-planar structural features, such as seams, folds, hems, layered regions, and combinations of different fabrics.} Among these structural features, seams are particularly common, because textile products are typically produced by transforming 2D fabrics into 3D structures through pattern cutting and sewing assembly. In addition, seams are often used to create forms and structural reinforcement in textile products. Seam placements frequently correspond to structurally or functionally meaningful regions, such as panel transitions or areas of motion, e.g., shoulder seams, underarm seams, and side seams in textile products. These regions often experience motion or strain, making them promising locations for embedding sensors or circuits.

\section{Related Work}

We structure our related work into four parts: (1) integrating functionalities into textiles; (2) material variability in fabrication, which has been studied across different fabrication domains; (3) robust fabrication concepts that address machine and fabrication uncertainties; and (4) digitization in fabrication, where toolpath generation is used to adapt designs to fabrication conditions.



\subsection{Integrating functionalities into textiles}
Smart textiles create opportunities to embed sensing~\cite{shao2025my}, actuation~\cite{zhang2025magtex}, wireless transmission~\cite{takahashi2022meander}, and interaction~\cite{lin2024tactex} into everyday soft materials. 
Prior work has integrated functionality into textiles through methods including knitting~\cite{li2025plug, zhang2025magtex}, weaving~\cite{sun2020weaving, lin2024tactex}, sewing~\cite{ibrahim2025sewing, yu2025seamfit}, serging~\cite{ibrahim2025serging, ruston2024seamsleeve}, lamination~\cite{zhou2023mocapose}, and embroidery~\cite{hamdanSketchStitchInteractiveEmbroidery2018}. Among these, machine embroidery is distinctive because it can programmatically place functional threads onto existing textile products with precise control over stitch placement and routing~\cite{hamdanSketchStitchInteractiveEmbroidery2018}, making embroidery promising for augmenting existing soft goods such as garments, accessories, braces, and home textiles~\cite{Khorsandi_Jones_Davoodnia_Lampen_Conrad_Etemad_Nabil_2023,goudswaard_fabriclick_2020}.
Prior work has used embroidery to create diverse functions, such as pressure sensing structures~\cite{pointnerEmbroideringResonantCircuits2025, luo2024adaptive}, gestural interaction interfaces~\cite{aigner2021texyz, yu2023skinergy}, haptic output~\cite{jiang2024embrogami, nabil2019seamless}, and textile speakers~\cite{preindl2020sonoflex, nabil2021soft}. 
 \rev{E-Sewing~\cite{ibrahim2025sewing} detailed stitch strategies for integrating components, terminating connections, and insulating conductive thread.} These works demonstrate the opportunity to directly augment textile goods. Building on this opportunity, we focus on the next step: enabling reliable direct technical embroidery on assembled soft goods by addressing seams as a common source of structural variation. 


\subsection{Material variability}

Digital fabrication research has highlighted the importance of understanding material variability, as variations in material properties can influence design decisions \cite{batra_convivial_2026}, fabrication parameters, and the reliability of fabricated structures. 

Prior work has explored computational techniques to automate or assist the process of detecting and characterizing material variability. These approaches include techniques such as computer vision, sensing~\cite{dogan2021sensicut}, and machine learning models that analyze material structures. For instance, computer vision techniques have been applied to classify defects in wooden structures from images, enabling more efficient inspection and assessment of material quality~\cite{Ehtisham_Qayyum_Camp_Plevris_Mir_Khan_Ahmad_2024}. \rev{Others have taken advantage of these variations for identification of objects \cite{dogan_structcode_2023}.} 

Beyond detecting and characterizing material variability, prior work has also used this information to guide fabrication decisions, such as selecting fabrication parameters or adjusting designs accordingly. \rev{Fabricaide \cite{fabricaide2021clean} addresses fabrication with existing leftover materials, where each sheet differs in available area, shape, and preexisting holes, and shows how these material-specific constraints can be captured and integrated into the design process to help users adapt their designs before fabrication. Similarly,} SensiCut integrates speckle sensing and deep learning to distinguish visually similar materials on laser cutters, and then uses the detected material properties to recommend fabrication parameters or design adjustments ~\cite{dogan2021sensicut}.

Inspired by these approaches, we bring the concept of material-aware fabrication to machine embroidery by developing a tool that detects textile variability and generates fabrication parameters and design adjustments.

\subsection{Robust design}

Fabrication processes inevitably introduce variations due to factors such as tool accuracy, material properties, and machine calibration. 
\rev{Adaptive fabrication processes address such variations by sensing process changes and making adjustments in real time; for example, high-end ZSK embroidery systems use computer vision to detect material changes (e.g., shrinkage) during stitching and make real-time adjustments. Similarly, \textit{Kerfmeter} \cite{katakura2023kerfmeter} is a calibration routine for laser cutters prior to cutting to guarantee precision.} The robust design methodology in mechanical engineering aims to make systems insensitive to variations introduced by such noise factors, allowing systems to remain functional despite inevitable variations in manufacturing or operation ~\cite{Arvidsson_Gremyr_2008}. Inspired by this concept, prior work in digital fabrication methods, KerfCanceler and SpringFit incorporate self-adjusting mechanisms that compensate for fabrication tolerances across different machines and fabrication conditions~\cite{roumen_kerf-canceling_2020,roumen_springfit_2019, roumen_portable_2023}. These works demonstrate how embedding tolerance-aware mechanisms directly into the design can expand the range of fabrication conditions under which systems remain functional. In this work, we extend this perspective to machine embroidery, exploring how robust design principles can improve the reliability of technical embroidery.

\subsection{Digitization in fabrication}

The role of computer-aided manufacturing (CAM) in material variation is essential in addressing the challenges and complexities that arise when working with diverse materials. In digital fabrication, path planning is often used to adapt designs to different manufacturing conditions. For example, in 3D printing, slicing converts 3D models into machine-executable tool paths while considering material properties, printer characteristics, and fabrication parameters, thereby improving fabrication stability and success rates. Feng et al. explore how to explicitly convert between different CAM workflows \cite{feng_cameleon_2024}. Machine embroidery similarly relies on digitization, which converts design graphics into stitch paths and machine parameters. Advanced commercial embroidery digitizing software such as Wilcom/Hatch allows users to select fabric types and automatically adjust stitch parameters (e.g., density or underlay), but these approaches assume homogeneous materials across the entire design. Rather than modifying the design itself, we explore how digitization can be used to adapt stitch paths to these structural variations, allowing the same design and functional threads to be reliably embroidered across different seam and material conditions.

\section{Mechanism design of \stitches{}}
The main contribution of this paper is the mechanism of \stitches{}. These are modifications to stitch patterns to allow crossing seams without causing defects or functional failure. In this section we lay out their main mechanism and the other types of stitches we introduce to facilitate successful use of \stitches{}. To understand the mechanism of these solutions, we first dive deeper into what causes problems of embroidering across seams in the first place.

\subsection{The problems of embroidering across seams}

\begin{figure}[h]
  \centering
  \includegraphics[width=\linewidth]{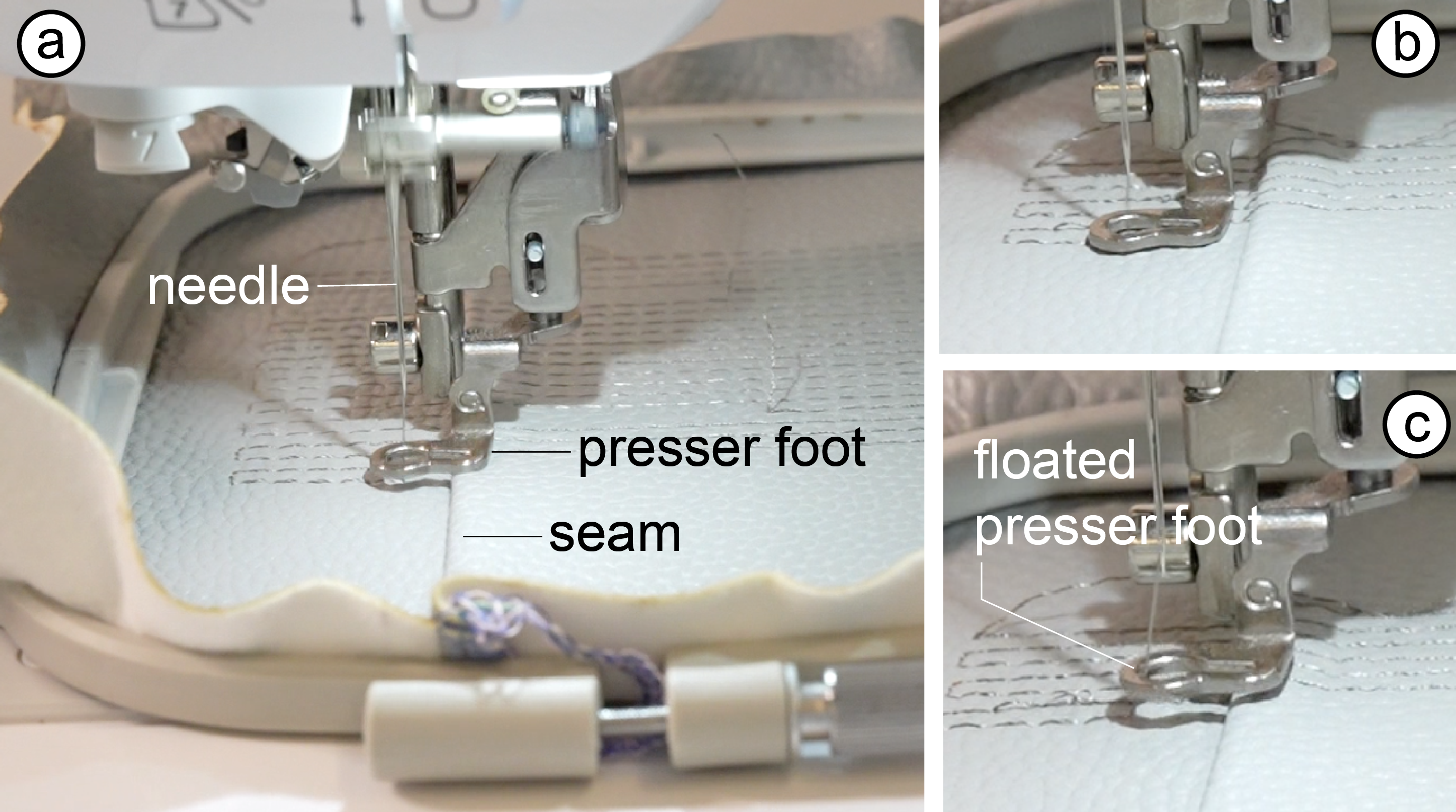}
  \caption{(a)~Side view of the embroidery machine. (b)~Presser foot fully pressed down on flat fabric. (c)~Presser foot lifted by the seam, resulting in unstable stitching conditions.}
  \label{fig:embroidery-machine}
\end{figure}

\begin{figure}[h]
  \centering
  \includegraphics[width=\linewidth]{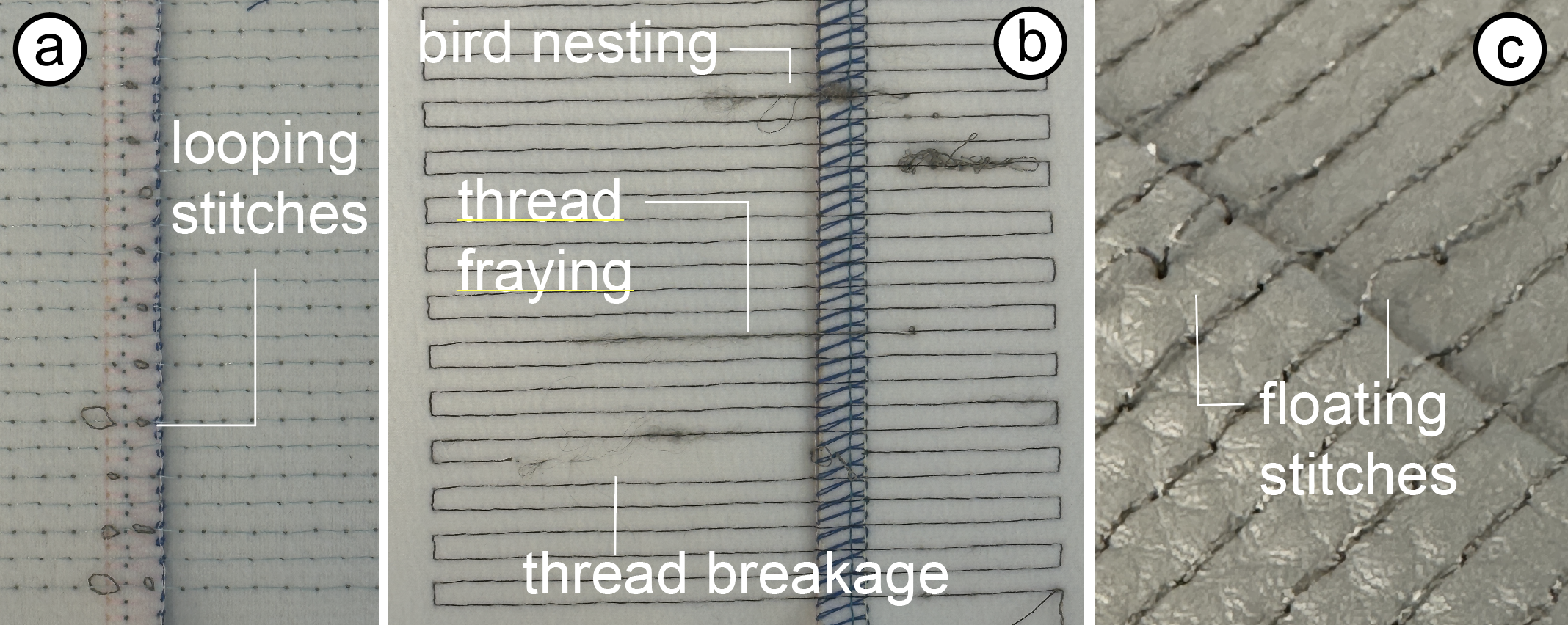}
  \caption{Typical problems caused by seams which lead to fabrication defects.}
  \label{fig:seam-problems}
\end{figure}

Home embroidery machines employ a spring-loaded “presser foot” (labeled in Figure~\ref{fig:embroidery-machine}) that lifts and descends in synchrony with the needle cycle, pressing the fabric against the needle plate (the platform that supports the fabric during embroidery) to stabilize it during stitch formation. When the presser foot encounters a seam, the increase in substrate thickness forces the presser foot upward against its spring. This produces three concurrent effects: (1)~the spring is more compressed, which increases the downward pressure on the fabric; (2)~this in turn increases friction between the presser foot and the fabric surface, impeding smooth hoop transport as the embroidery frame continues its programmed XY motion, thus introducing positional errors; and (3)~the raised presser foot fails to fully press the fabric against the needle plate, destabilizing the substrate and having a floating needle which needs to press further down to reach the substrate. Acting together, these effects alter the tension and friction conditions between the thread and fabric during stitch formation, giving rise to failure modes including thread breakage, thread fraying, bird nesting, and pattern distortion highlighted in Figure~\ref{fig:seam-problems}. 

\begin{figure}[h]
  \centering
  \includegraphics[width=\linewidth]{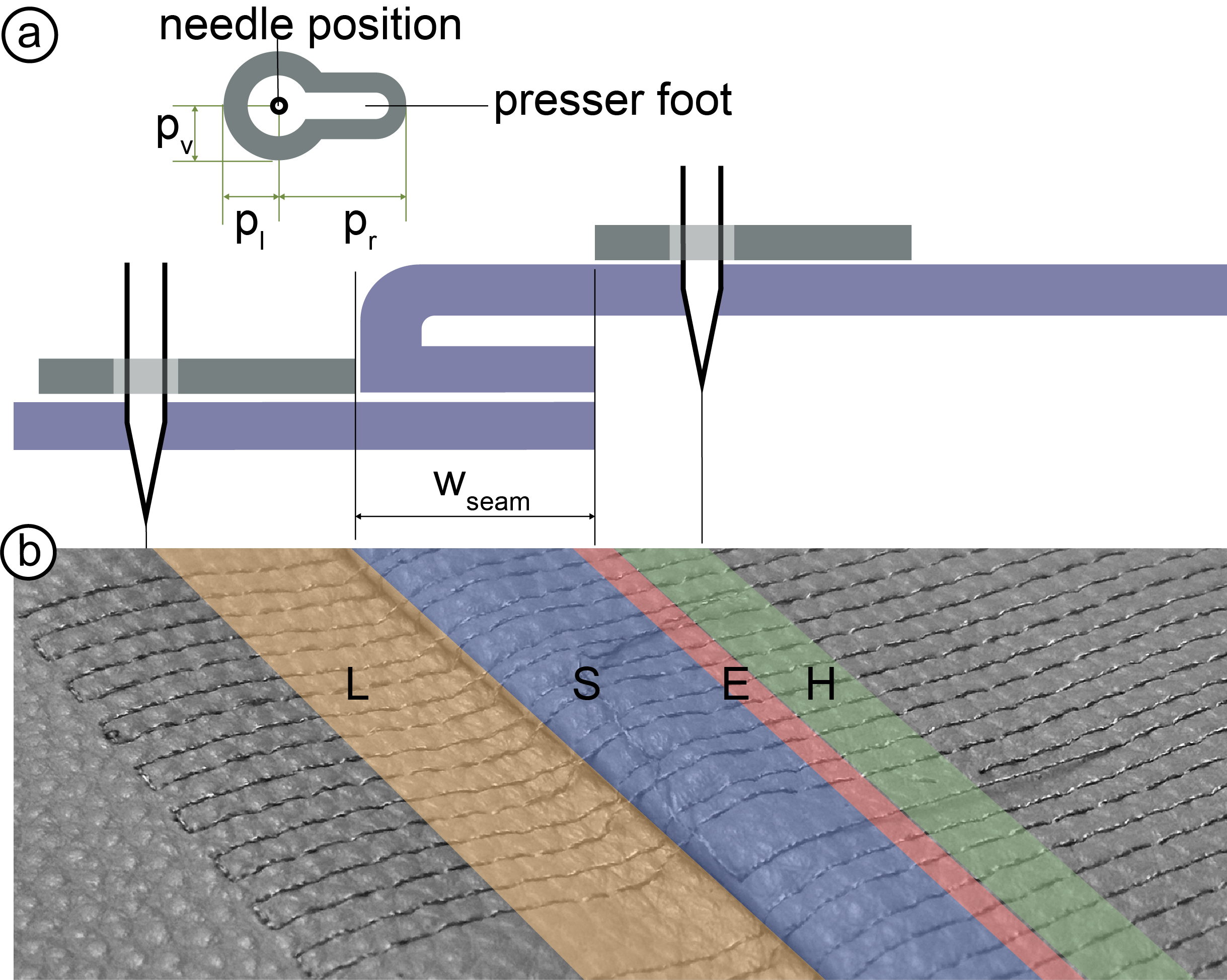}

  \caption{The risky regions in which technical embroidery may fail due to a seam. (a) The shape and position of the presser foot determine the width of \rev{L and H}. (b) A side-view of the folded seam during embroidery.
}
  \label{fig:seam-regions}
\end{figure}

Based on this problem definition, there are four risky regions around the seam.
When needle points are placed within these regions, the embroidery performance worsens, as illustrated in Figure~\ref{fig:seam-regions}. \rev{S is the overlapping seam area. Besides the seam itself, L (low) and H (high) are the areas on the lower and higher sides of the seam, respectively. E (edge) is the line where S meets H. L’s risk is caused by the thicker S preventing the presser from fully pressing down. H’s risk is caused by the hollow space underneath. E’s risk is caused by the cliff-like thickness change between S and H.} On the seam (S) itself, it is risky to embroider from across the seam, but a path that lies entirely on the seam, all needle points experience consistent substrate conditions, allowing thread tension to reach a stable state, (technical) embroidery there only works if the path is fully in the S region. The problems of embroidery in these regions are further amplified when the belt tension of the machine is either over- or under-tensioned. Compared to decorative embroidery, technical embroidery operates under tighter tolerances, making it more sensitive to such variations, as validated in our technical evaluation.

\subsection{Our proposed solution: \stitches{}}

\begin{figure}[h]
   \centering
   \includegraphics[width=\linewidth]{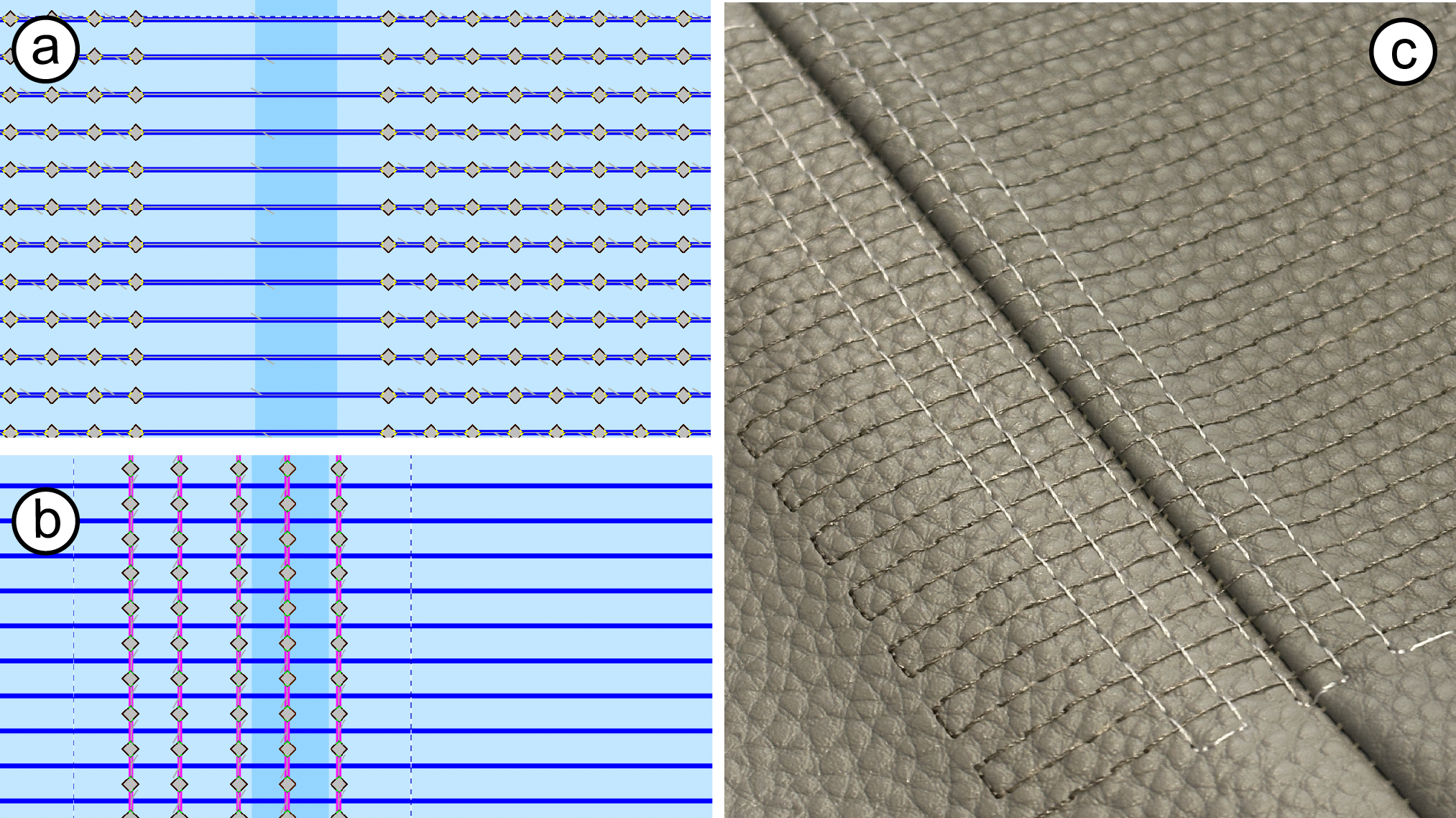}
   \caption{Our \stitches{} address problems caused by seams. (a) placing the \stitches{} across the seam and (b) adding alignment stitches parallel to and on the seam to keep the \stitches{} in place.}
   \label{fig:jumpstitches}
 \end{figure}

We introduce what we call “\stitches{}” visualized in Figure~\ref{fig:jumpstitches}. The key idea is to remove the needle points within the risky regions and instead make a larger jump on the fabric across the seam. This jump circumvents the problems outlined in Figure~\ref{fig:seam-problems}, but it produces a relatively loose conductive thread. To keep it in place, we add a series of alignment stitches parallel to the seam on either side. 

The minimum jump length is determined by the geometric relationship between the presser foot, embroidery path and the seam and spans all three regions of Figure~\ref{fig:seam-regions}. Specifically, the jump must be sufficiently large to ensure that the presser foot does not intersect or interact with the seam along its motion path, thereby fully bypassing the high-risk interaction region. As shown in Figure~\ref{fig:jumpstitches}a, these long jumps produce somewhat loose functional thread along the seam. To counter this effect, we insert alignment supporting stitches in the seam region (shown in  Figure~\ref{fig:jumpstitches}b) and right before the jump. These stitches are formed parallel to the seam and thus do not suffer from the instabilities of interacting with the seam.

As demonstrated in our technical evaluation, \stitches{} address the problem of defects and drop in resistance across seams, however, their geometric nature also explicitly introduces some problems: (1)~by removing needle points from the original design there is a risk that what we refer to as shape-defining needle points (needle points that define the shape of the pattern, e.g. points on the turning points of the shape) are removed, breaking the circuitry and (2)~when multiple seams intersect the material can locally deform non-linearly which is hard to predict, resulting in failing \stitches{}. We propose two additional interventions to overcome these limitations.

 \begin{figure}[h]
   \centering
   \includegraphics[width=\linewidth]{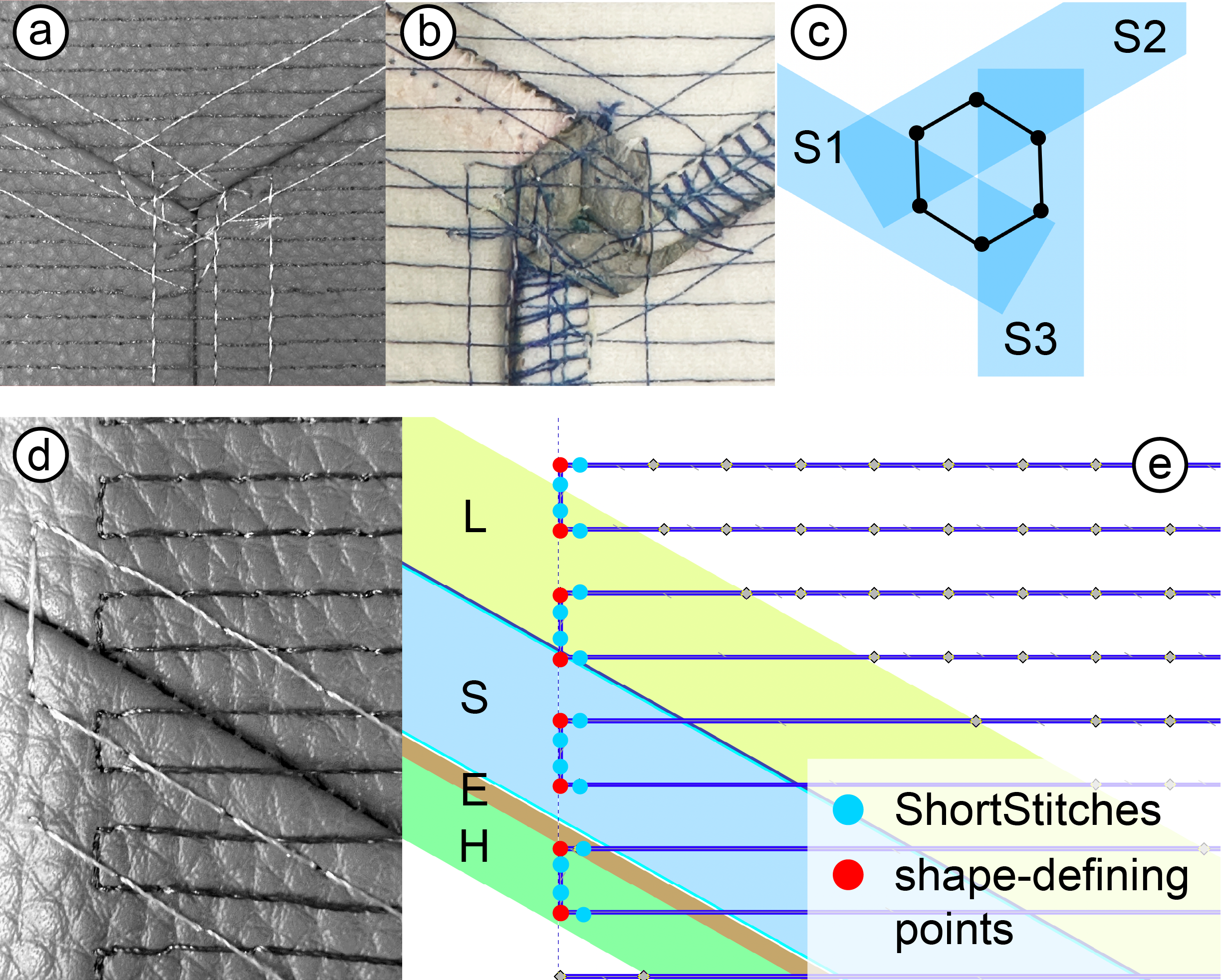}
    \caption{Two special cases of support stitches to overcome limitations of \stitches{}. (a,~b,~c)~At three-way seam intersections, we generate \underlaystitches{} that flatten the overlapping fabric layers: (a)~front side of the fabric, (b)~back side, and (c)~the \underlaystitches{} schematic diagram. (d,~e)~When a shape-defining point falls within a risky region (L, H, or E, as defined; the seam itself is labeled S): (d)~the \shortstitches{} solution as embroidered on fabric, and (e)~the corresponding \shortstitches{} schematic diagram.}
   \label{fig:support-stitches}
 \end{figure}

\paragraph{\shortstitches{} for shape-defining points.} When shape-defining points lie in the risky regions (S, L, H, and E), as illustrated in Figure~\ref{fig:support-stitches}e, we introduce localized \shortstitches{} which have a reduced stitch length of 0.5 mm. By reducing the in-plane (XY) step size between consecutive needle penetrations, the presser foot experiences smaller incremental vertical (Z) displacements when encountering thickness changes. This effectively transforms an abrupt height discontinuity into a sequence of gradual transitions, allowing the presser foot to traverse the seam more smoothly (sample results shown in Figure~\ref{fig:support-stitches}d).

\paragraph{\underlaystitches{} at Seam Intersections} 
At multi-way seam intersections (e.g., three-way, four-way, or higher-order seams), accumulated thickness and overlapping fabric layers introduce significant irregularities that increase the risk of stitching failures. To mitigate this, we define a minimal boundary region around the seam intersection (Figure~\ref{fig:support-stitches}c) and generate a stitch-down pattern within this region, which we refer to as \underlaystitches{}. \underlaystitches{} flatten and stabilize the intersecting fabric layers, providing consistent support and reducing local interference as the embroidery path traverses the junction (An example is shown in Figure~\ref{fig:support-stitches}a and b).

\section{\name{}}
\begin{figure}[h]
  \centering
  \includegraphics[width=\linewidth]{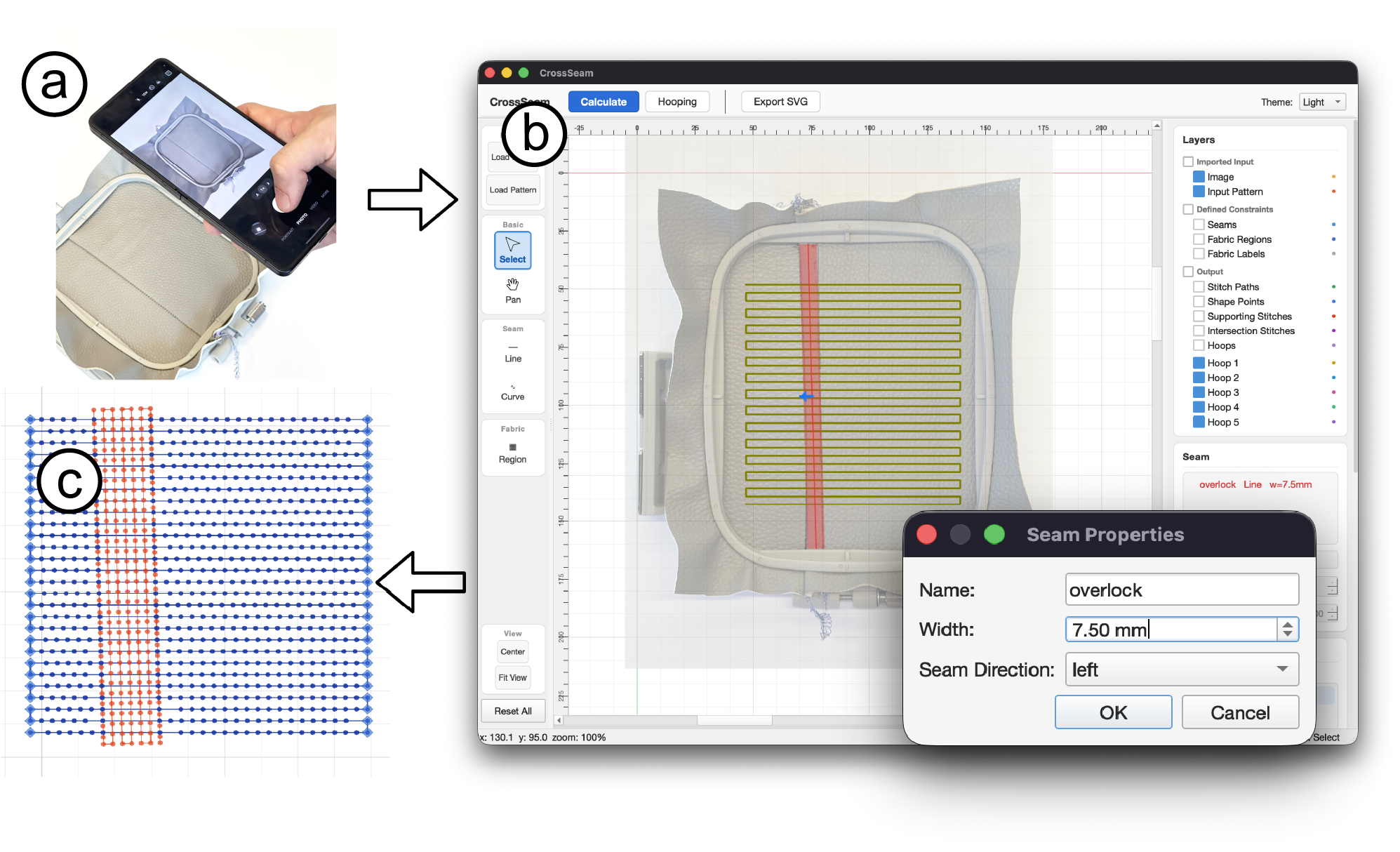}
  \caption{Workflow of \name{}. (a)~The user takes a photograph of the hooped fabric. (b)~The hoop image and stitch pattern are imported into \name{}, where the user aligns the pattern to the desired location and defines seams using the photograph as a reference. (c)~\name{} generates the modified stitch plan with JumpStitches.}
  \label{fig:workflow}
\end{figure}

We build a simple software tool \rev{\name{}} to let users take advantage of our \stitches{}. As shown in Figure~\ref{fig:workflow}, users hoop their material, take a picture of the hooped material with seams, and our tool converts the original stitch pattern into a pattern with \stitches{}, which the user then embroiders on the material.

The core functionality is simple: the software shows the image of the hoop, with the stitch pattern overlaid on top of it. Users use the seam tool to specify start and end points, as well as the direction of each seam. They can use Bézier curves for complex curved seams. Once they confirm the seam placement, the software automatically adjusts the needle points and exports the modified pattern as SVG, ready to be sent to the embroidery machine. 

Based on the embroidery path, user-defined seams, and the relative position between the presser foot and the seam, \name{} partitions the space along the path into four region types: S, L, H, and E. As a first step, \name{} checks whether the start or end points of the path fall within any of these regions. If either endpoint lies in a risk region, repeated local puncturing at that location may lead to defects. In such cases, the system issues a warning to inform users of the fabrication risk associated with this path. Next, \name{} identifies shape-defining points along the path, i.e., points that define the geometry of the pattern. It then analyzes whether two consecutive shape-defining points lie within the same risk region. In this case, the system provides high-level adjustment suggestions, including translation, scaling, rotation, or local path adjustment. If the user selects local path adjustment, the system moves the needle points at the nearest location outside of the risky regions and asks the user to confirm this change. If two consecutive shape-defining points do not lie within the same risk region, \name{} considers the segment suitable for stitch-level optimization. \name{} inserts \shortstitches{} around those needle points. After processing shape-defining segments, the system analyzes the remaining path segments that cross seams. For segments that are not geometry-critical, the system applies \stitches{} to address the seam problems. Finally, for all remaining segments, the system generates needle points based on the user-specified stitch length parameter, completing the overall stitch plan.

\begin{figure}[h]
  \centering
  \includegraphics[width=\linewidth]{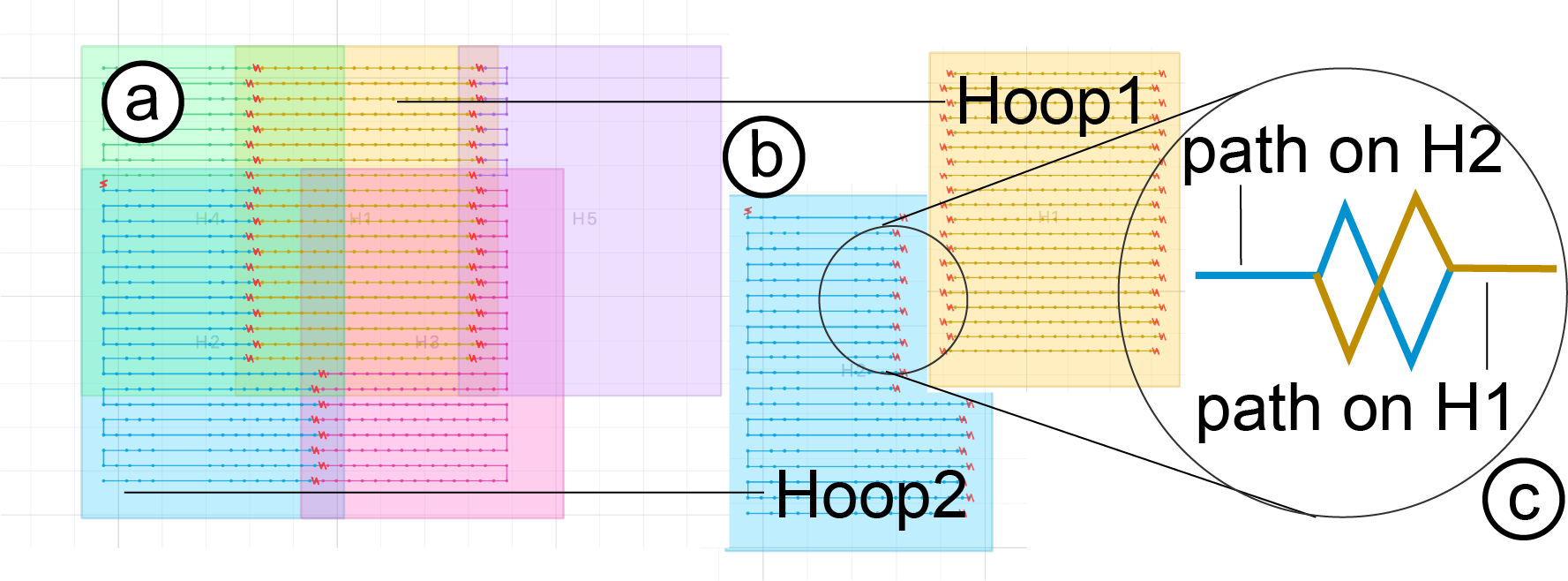}
  \caption{ Multi-hooping (a)~colored hoops are overlaid on the stitch pattern in the editor (b)~when two hoops overlap across a circuit, \name{} inserts \jointstitches{} (c)~close-up of a pair of default \jointstitches{}.}

  \label{fig:multihooping}
\end{figure}

\paragraph{Multi-hooping: extending to larger fabrics} 
To really work on everyday textiles as well as with complex stitch patterns, \name{} supports \textit{multi-hooping}. This is a process where stitch patterns are larger than the size of the embroidery hoop. \name{} allows users to define a customized size of the hoop and split the stitch patterns into sub-patterns that could fit into each hoop size. We adopt a similar hooping algorithm presented in \textit{OriStitch}~\cite{chang_oristitch_2025} to greedily place hoops such that they can cover the most remaining calculated stitch points iteratively for each hoop such as Figure~\ref{fig:multihooping}. In addition, customizable joint stitches are appended to connect interrupted paths as shown in Figure~\ref{fig:multihooping}. The default joints are zigzag lock stitches on top of one another; for conductive threads with exterior coating, this still maintains continuity. Also, users are allowed to upload customized connector pattern for different needs.


\paragraph{Different fabric settings} 
Seams connect two fabrics to one another, in most cases these are the same material, but it is possible to use seams to connect two different materials too. \name{} segments the fabric along the seams and prompts users to specify material properties of each segment. \name{} accommodates for different materials by adjusting the distance between needle points, it also updates the length of the \stitches{} for seams that connect to each material as they are proportional to the thickness of the two materials that are seamed together. 


\begin{figure}[h]
  \centering
  \includegraphics[width=0.5\linewidth]{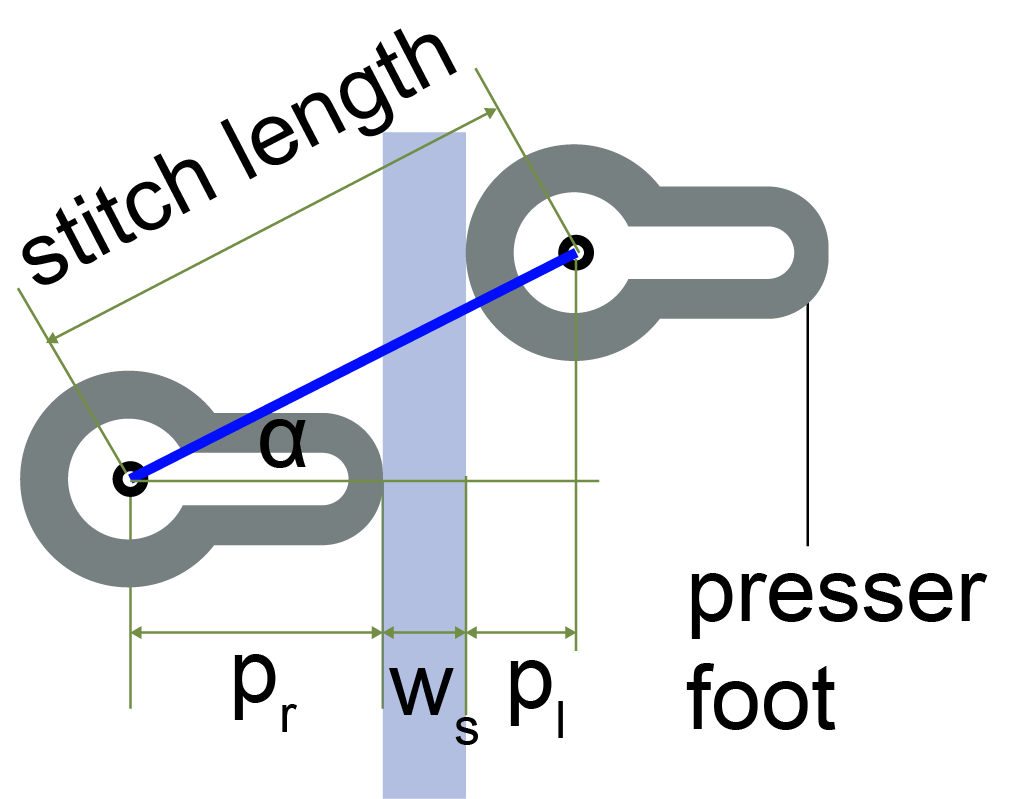}
  \caption{\name{} restriction minimum jump length is defined by seam and presser foot.} 
  \label{fig:resolution}
\end{figure}

\paragraph{\stitches{} resolution limit} 
The \stitches{} work for most scenarios, however the minimum jump length is physically restricted. The minimum jump stitch length \name{} supports is defined by the presser foot dimensions $P_r$ and $P_l$, the seam width $W_s$, and the angle $\alpha$ between the thread path and the pressor foot. As shown in Figure~\ref{fig:resolution}, $P_r$ and $P_l$ represent the minimum distances from the seam at which the presser foot can rest flat on the left and right sides, respectively. The minimum jump stitch length is given by:
\begin{equation}
L_{\min} = \frac{P_r + W_s + P_l}{\cos(\alpha)}
\end{equation}

\section{Implementation}

\name{} is implemented as a standalone desktop application using Python and the PySide6 (Qt) framework. It mostly converts SVG stitchplans and thus takes both as input and output SVG which can be sent straight to the digitizer (driver) of the embroidery machine.

The front-end provides a millimeter-based canvas with pan, zoom, and direct manipulation of all design elements. Users import an embroidery design as an SVG file and, optionally, a photograph of the physical fabric for visual alignment. Two seam annotation tools are provided: a straight-line tool for simple seams and a spline tool for curved seams. Each seam carries editable properties, including width, left/right buffer distances, and a normal direction indicating the fold orientation. Seams automatically partition the fabric into regions; users may assign per-region stitch parameters such as stitch length through the property panel.

\section{Technical evaluation}
To evaluate the technical functionality of \stitches{}, we conducted a series of Technical evaluations. Each evaluation studies a material or machine property which we evaluate under three conditions: a plain patch of fabric, a seamed fabric, and the seamed fabric with \stitches{}. We measure detected defects, the resistance of the used conductive thread, and positional error.

\subsection{Materials}
\begin{figure}[h]
  \centering
  \includegraphics[width=\linewidth]{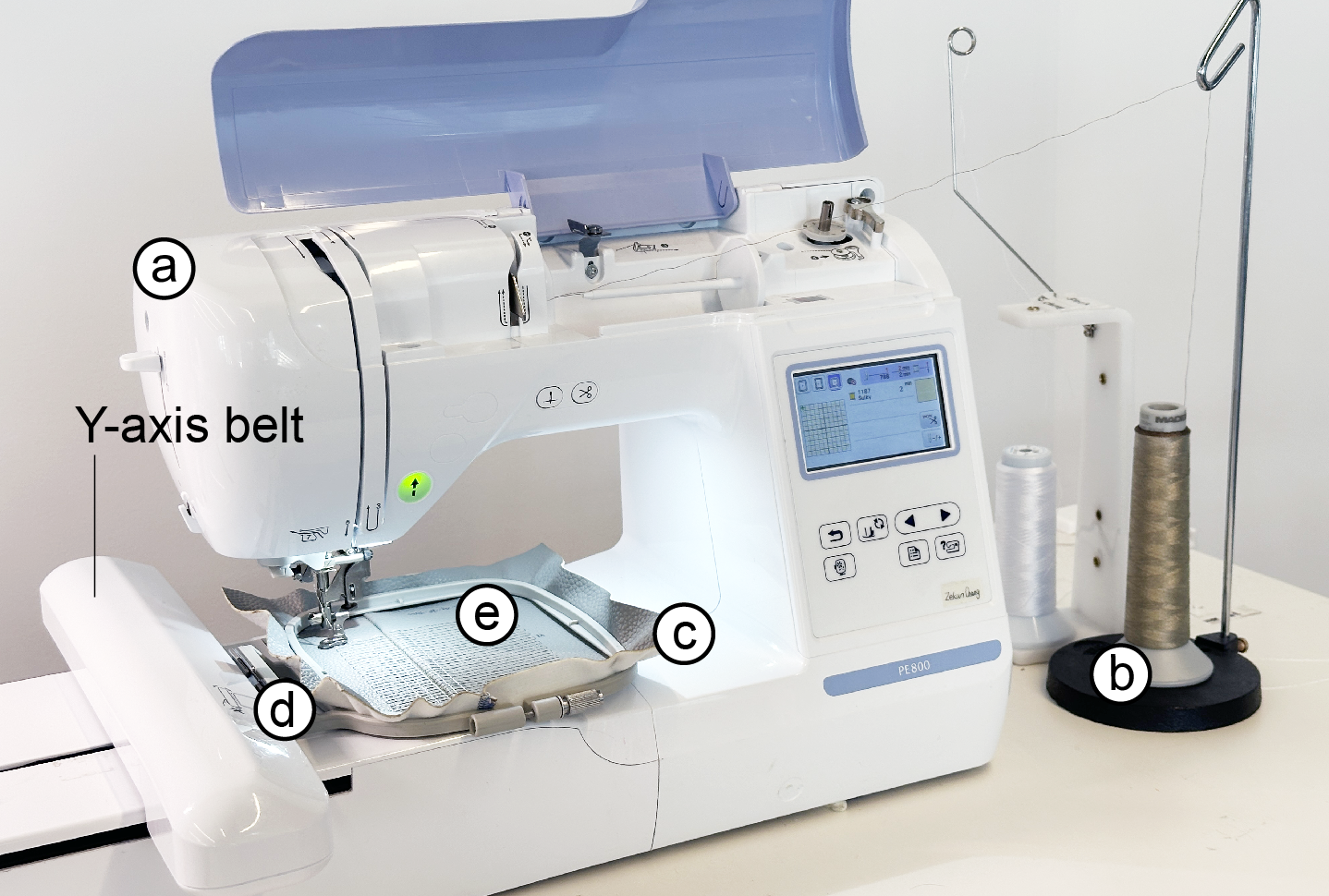}
  \caption{We conducted experiments using (a)~the Brother PE800 embroidery machine, (b)~with Madeira HC~40 thread and (c)~PU leather as substrate hooped in a (d)~100 x 100~mm hoop, and (e)~embroider a simple capacitive sensor.}
  \label{fig:materials}
\end{figure}

Figure~\ref{fig:materials} shows the set-up of our evaluation. All embroidery was performed on a Brother PE800, a typical home-use embroidery machine. We ran it at its lowest speed setting (350spm), using Madeira HC~40 conductive thread as the top thread and pre-wound 60wt polyester as the bobbin thread, with a universal embroidery needle (75/11) as recommended by the thread manufacturer. Before running the evaluation we took the machine to a maintainer who adjusted belt tension, oiled and checked the relevant mechanisms, to make sure it was operating in its best possible conditions. We opted to measure conductive technical embroidery as a particularly challenging use case and one for which it is relatively easy to measure performance. The thread tension dial was set to 4, a value empirically determined to produce consistent, defect-free embroidery on plain fabric. The substrate was PU leather, with 6~mm seams constructed using a Juki MO-654DE serger in a standard overlocked setting. 

We used a 100 × 100~mm hoop and embroidered a pattern consisting of 28 parallel rows of running stitches (stitch length: 4~mm), each 90~mm long. We repeated each sample 10 times.

We embroider a simple capacitive proximity sensor composed of a connected grid of straight lines. We chose this as a lower bound on complexity, which we use for most experiments, we follow up with a more complex pattern to set the higher bound in our last experiment. 

\subsection{Metrics}
To measure the quality of embroidery, we counted defects (shown in Figure~\ref{fig:seam-problems}) through visual inspection and classified them into three categories based on severity:

\paragraph{Minor defects} are aesthetic in nature and do not immediately compromise functionality, but may accumulate over time and eventually lead to failure: \textit{(a)~looping stitches}, where excess thread forms visible loops on the fabric backside; \textit{(b)~floated stitches}, where stitches arch above the fabric surface; and \textit{(c)~thread fraying}, where the thread shows visible fiber separation along its length, indicating localized wear or damage to the conductive coating.

\paragraph{Functional defects} directly compromise the intended functionality of the embroidered circuit: \textit{(d)~thread breakage}, where the thread snaps during stitching, breaking electrical continuity; \textit{(e)~pattern distortion}, where the stitched geometry deviates from the intended path, altering circuit layout; and \textit{(f)~support stitch penetration}, where supporting stitches penetrate conductive traces, potentially causing short circuits or damaging the conductive thread.

\paragraph{Process failures} indicate that the embroidery process itself has become unstable: \textit{(g)~bird nesting}, where thread accumulates and tangles underneath the fabric, typically requiring the process to be paused and manually cleared; and \textit{(h)~emergency stop}, where an emergency stop was manually triggered when abnormal machine behavior was observed (e.g., sharp increases in operational noise).

While minor defects do not immediately impact functionality, they signal suboptimal stitching conditions and tend to precede more severe failures in prolonged embroidery runs.

Conductive thread allows us to directly measure embroidery performance via electrical resistance using a multimeter. We measured one visually defect-free trace per sample, defined as a trace with no observable mechanical failure such as thread breakage or bird nesting. The resistance in traces without visible defects can still be affected by the embroidery process, as the conductive coating of the Madeira HC~40 thread tends to get damaged under suboptimal conditions. Measuring traces with visible defects would be uninformative, as their resistance is likely infinite due to broken continuity.

\subsection{Simple seams: \stitches{} eliminate defects from 70\% (baseline)}
\begin{figure}[h]
  \centering
  \includegraphics[width=\linewidth]{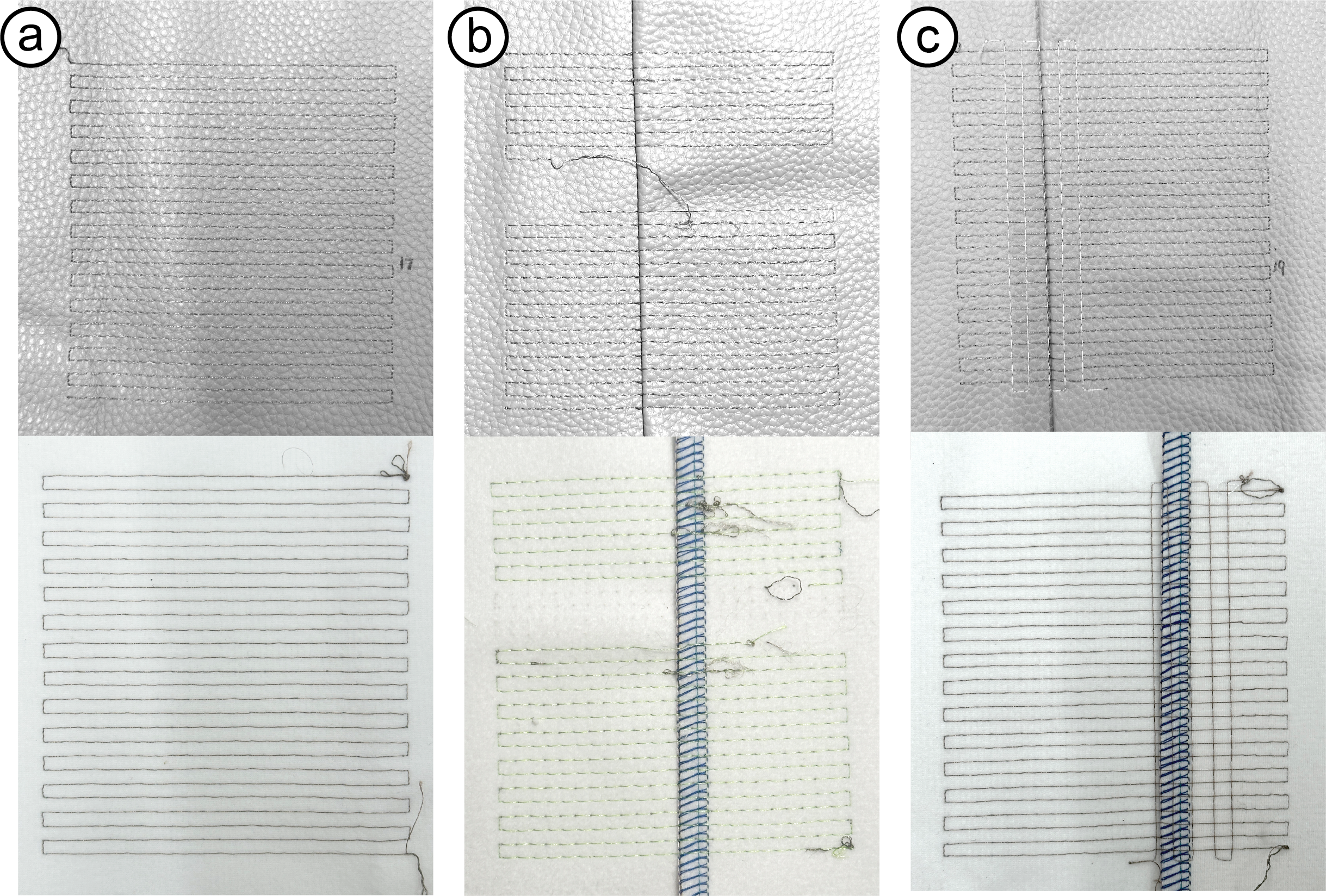}
  \caption{\rev{Representative results of the same embroidery pattern under three conditions: (a) plain fabric with traditional digitization, serving as the baseline; (b) seamed fabric with traditional digitization; and (c) seamed fabric with \stitches{}. Each column shows the front (top) and reverse (bottom) sides of a sample.}
}
  \label{fig:eval1}
\end{figure}

Our first evaluation is the basic effect of the seam on embroidery quality and how this changes with \stitches{}. As shown in Figure~\ref{fig:eval1}, our test set consisted of 10 samples of plain fabric, 10 samples with a seam and traditional digitization, and 10 samples with a seam and \stitches{}. 

\begin{figure}[b]
  \centering
  \includegraphics[width=\linewidth]{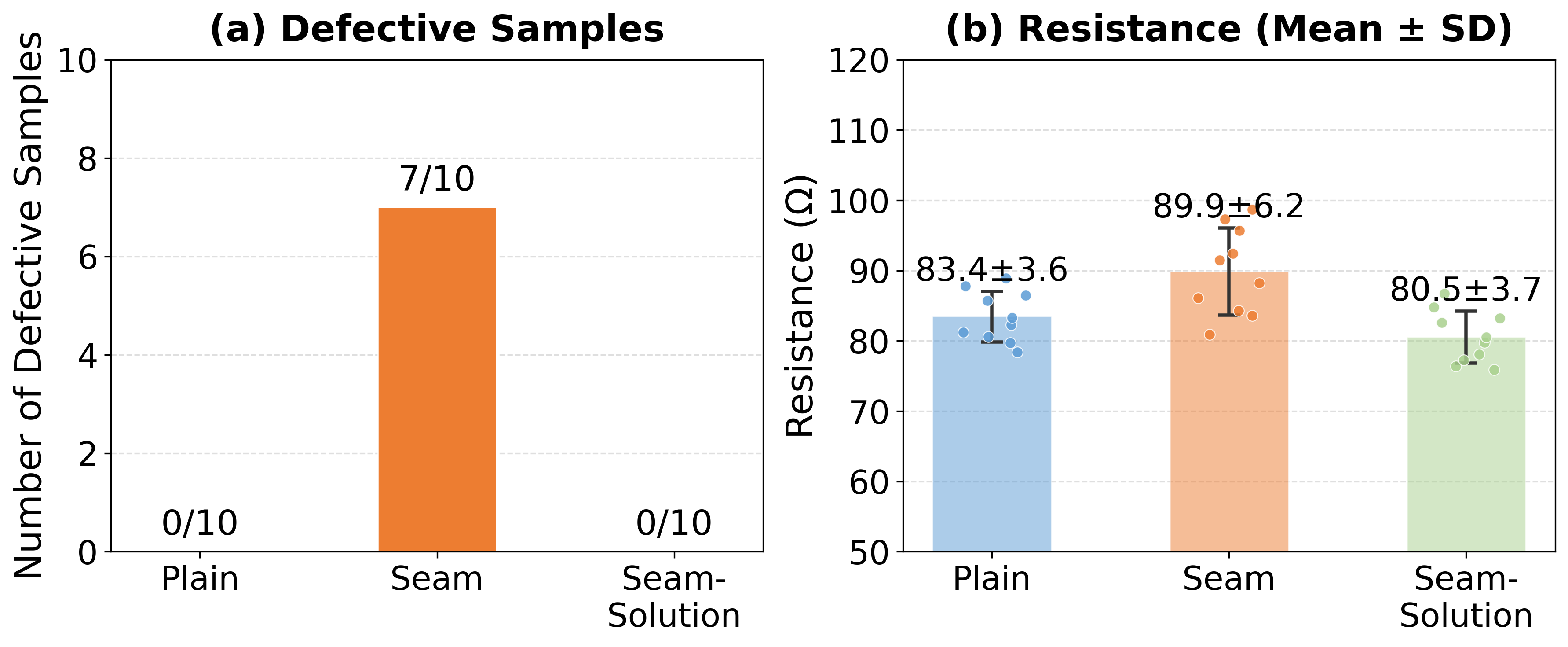}
  \caption{Embroidery quality and resistance. (a) Defective samples per construction method. (b) Resistances of individual measurements.}
  \label{fig:eval_wc}
\end{figure}

\subsubsection{Results}
As expected, plain fabric samples showed no defects across all 10 samples. Seam samples exhibited significant embroidery defects: 7 out of 10 samples showed defects and all as \textit{Looping Stitches}. The seam with \stitches{} showed zero defects across all samples, matching the defect-free performance of plain fabric. Electrical resistance remained within a comparable range across all three constructions albeit with elevated standard deviation (6.2 vs 3.7) in the seam condition, indicating small reduction in reliability of sensor performance.

\subsubsection{Conclusion}
We conclude that \stitches{} significantly reduce defects compared to the seam condition. We also show the problematic effect of seams in the first place, causing defects in 70\% of the samples and reducing the reliability of conductive threads. Our solution evidently overcomes both of these problems. 

\subsection{Belt tension: variations of belt tension increase defects, but \textit{not} with \stitches{}}
\begin{figure}[h]
  \centering
  \includegraphics[width=\linewidth]{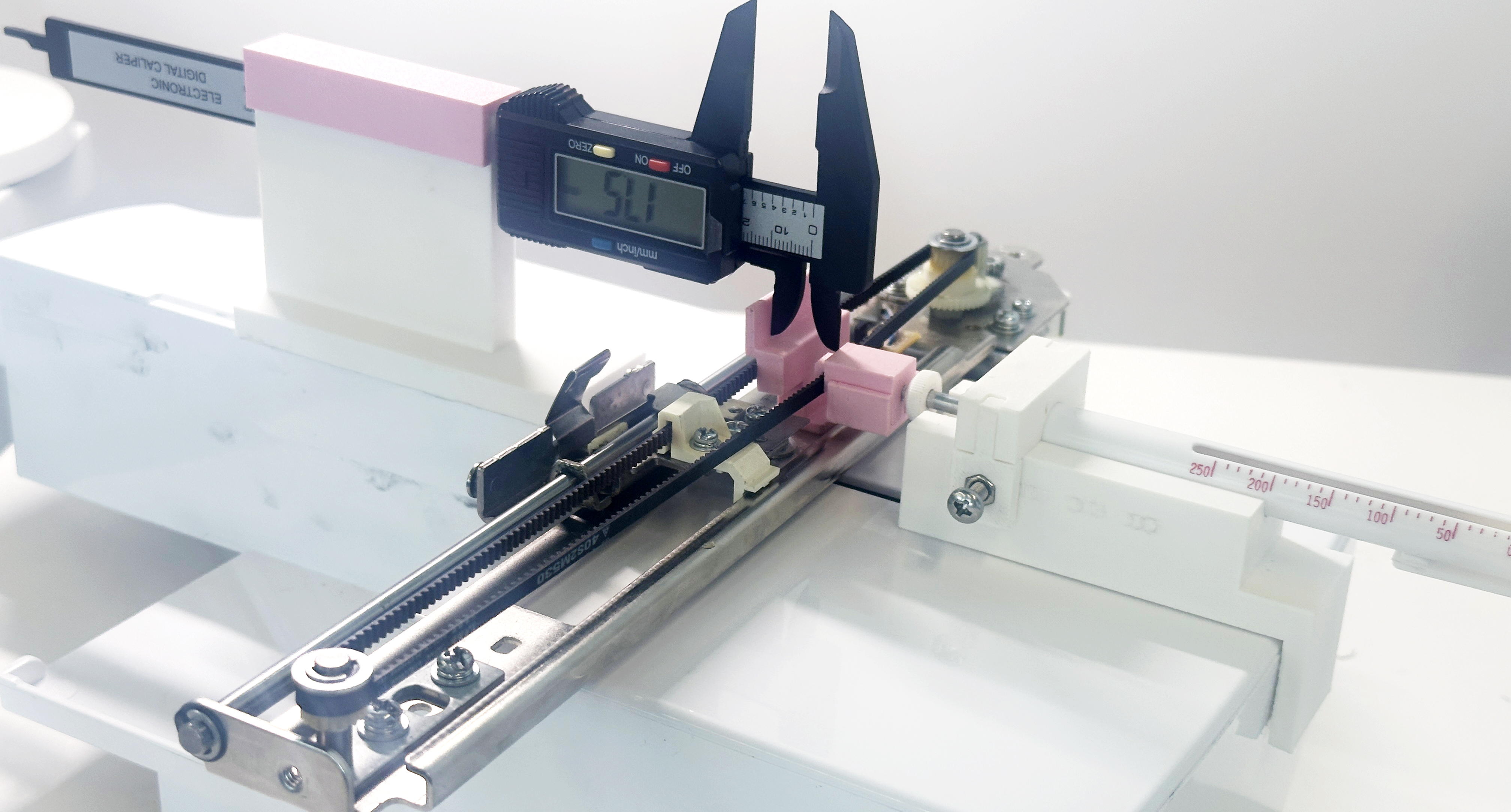}
  \caption{We used a force gauge and instrumented the machine to “de”-calibrate the Y-axis belt. We aligned the seams with this axis to maximize effects.}
  \label{fig:belt-tension}
\end{figure}

\rev{In practice, embroidery machines operate under different belt tensions (different from the thread tension) while remaining reliable on plain fabrics. A well-calibrated belt tension (e.g., after getting serviced) mitigates embroidery errors on seamed fabrics. However, seamed fabrics and technical embroidery threads are more sensitive to belt tensions; a machine state that produces no failures on plain fabric can still cause failures across seams. We therefore evaluated different tension conditions to test whether our method remains robust across a broader range of practical machine states.}
To evaluate whether \stitches{} support variations in belt tension, we conducted 2 more evaluation series in which we intentionally increased and reduced belt tension in the Y-axis from its well-calibrated state (the previous evaluation). We instrumented the machine to mount calipers and a force gauge (Figure~\ref{fig:belt-tension}). The three tension states, over-tensioned, well-calibrated, and under-tensioned, corresponded to deflection values of 13.2~mm, 15.5~mm, and 18.0~mm respectively, under a constant applied force of 0.49 N.

\begin{figure}[b]
  \centering
  \includegraphics[width=\linewidth]{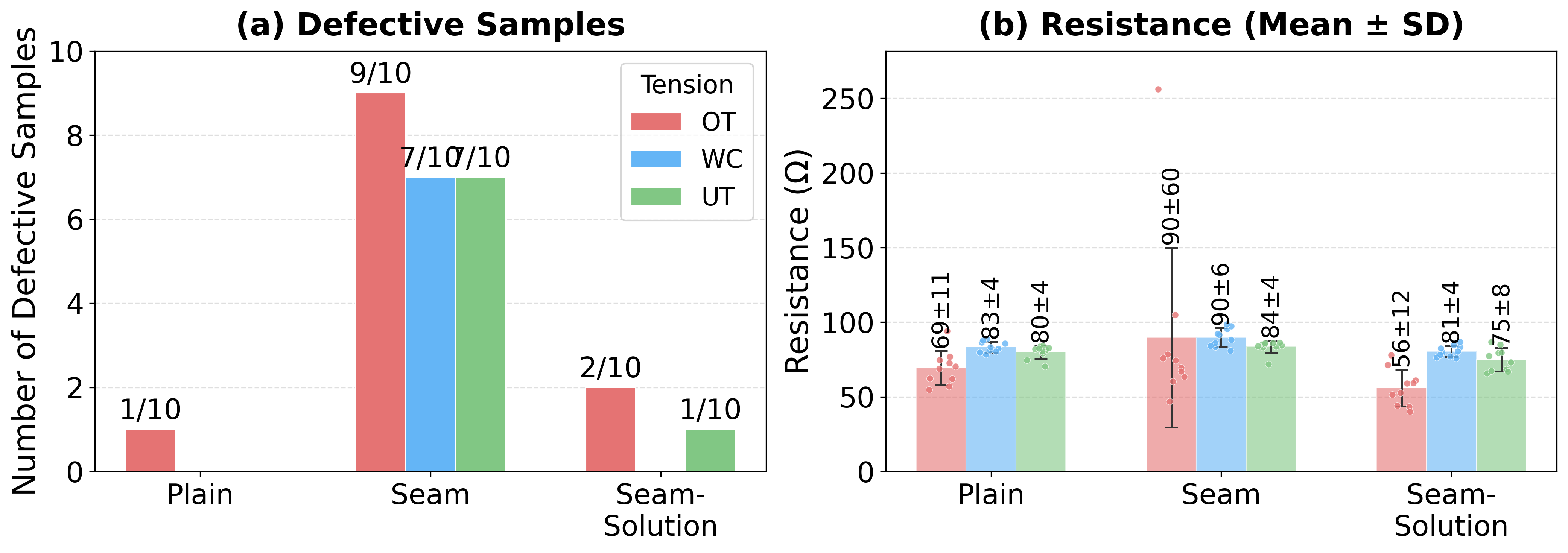}
  \caption{Comparison of embroidery quality across belt-tension conditions for the horizontal pattern. (a)~Number of defective samples out of 10 for each construction method under over-tension (OT), well-calibrated (WC), and under-tension (UT) conditions. (b)~Resistance measurements (mean ± SD) with individual data points.}
  \label{fig:eval-tension}
\end{figure}

\subsubsection{Results}
\begin{figure}[h]
  \centering
  \includegraphics[width=\linewidth]{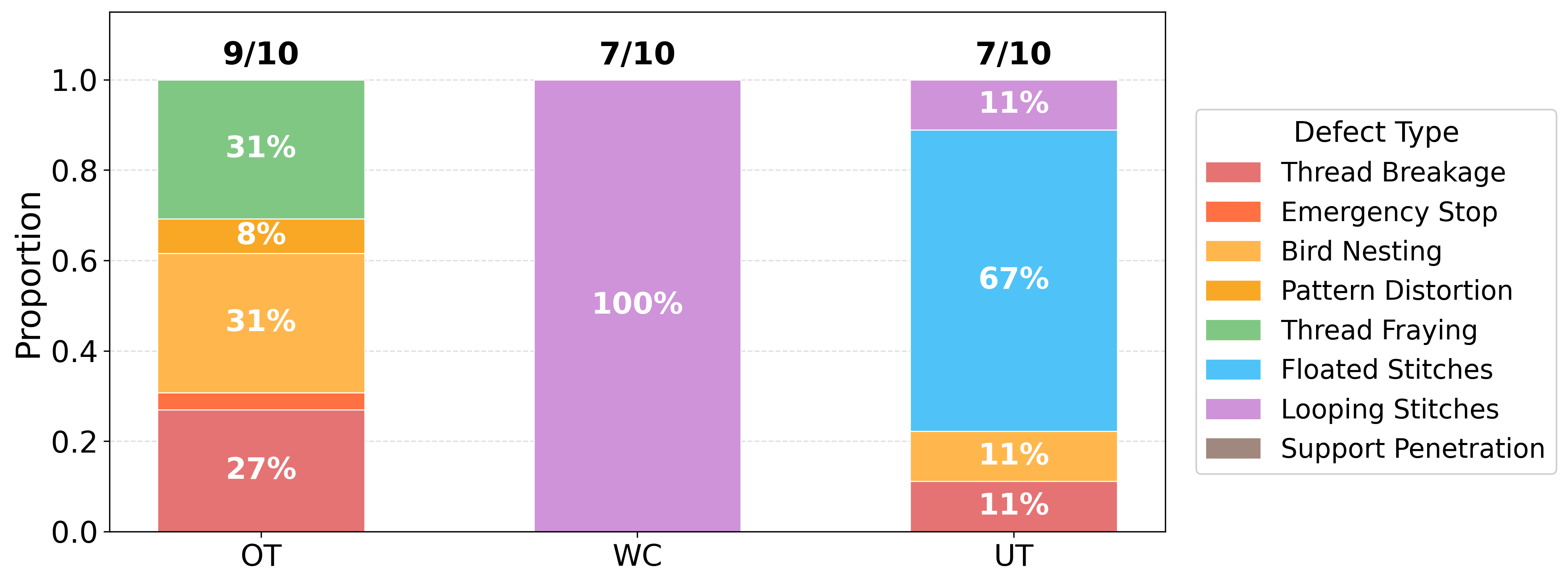}
  \caption{Proportion of defect types among defective seam samples per tension condition (horizontal pattern, n=10).}
  \label{fig:defect-types-by-tension}
\end{figure}

Figure~\ref{fig:eval-tension} shows the results of the variation in tension. Both over-tension and under-tension cause additional defects compared to the well-calibrated condition. While there are some defects in the \stitches{} condition (2 samples when over-tensioned, 1 when under-tensioned), they are primarily aesthetic (bird nesting, floated stitches, and supporting thread penetration). In the seam condition, over-tensioning leads to a sharp increase in defects (9/10 samples), with a wide variety of defect types including thread breakage, thread fraying, bird nesting, and pattern distortion. Under-tensioning produces the same number of defects as the well-calibrated condition (7/10), but the dominant defect type shifts from looping stitches to floated stitches. Figure~\ref{fig:defect-types-by-tension} shows how the type of defects changed when varying tension, from primarily floated stitches when under-tensioned to a variety of structural defects when over-tensioned.

The resistance again showed no significant difference in the \stitches{} condition. In the seam condition, we observe a large increase in standard deviation when over-tensioned, indicating further reduction in performance reliability under that condition.

\subsubsection{Conclusion}
The main conclusion is that variations in tension cause problems for technical embroidery and \stitches{} largely overcome this. The remaining defects were only aesthetic in nature and it is worth pointing out that even in the plain condition these defects showed up. 

\begin{figure}[h]
  \centering
  \includegraphics[width=\linewidth]{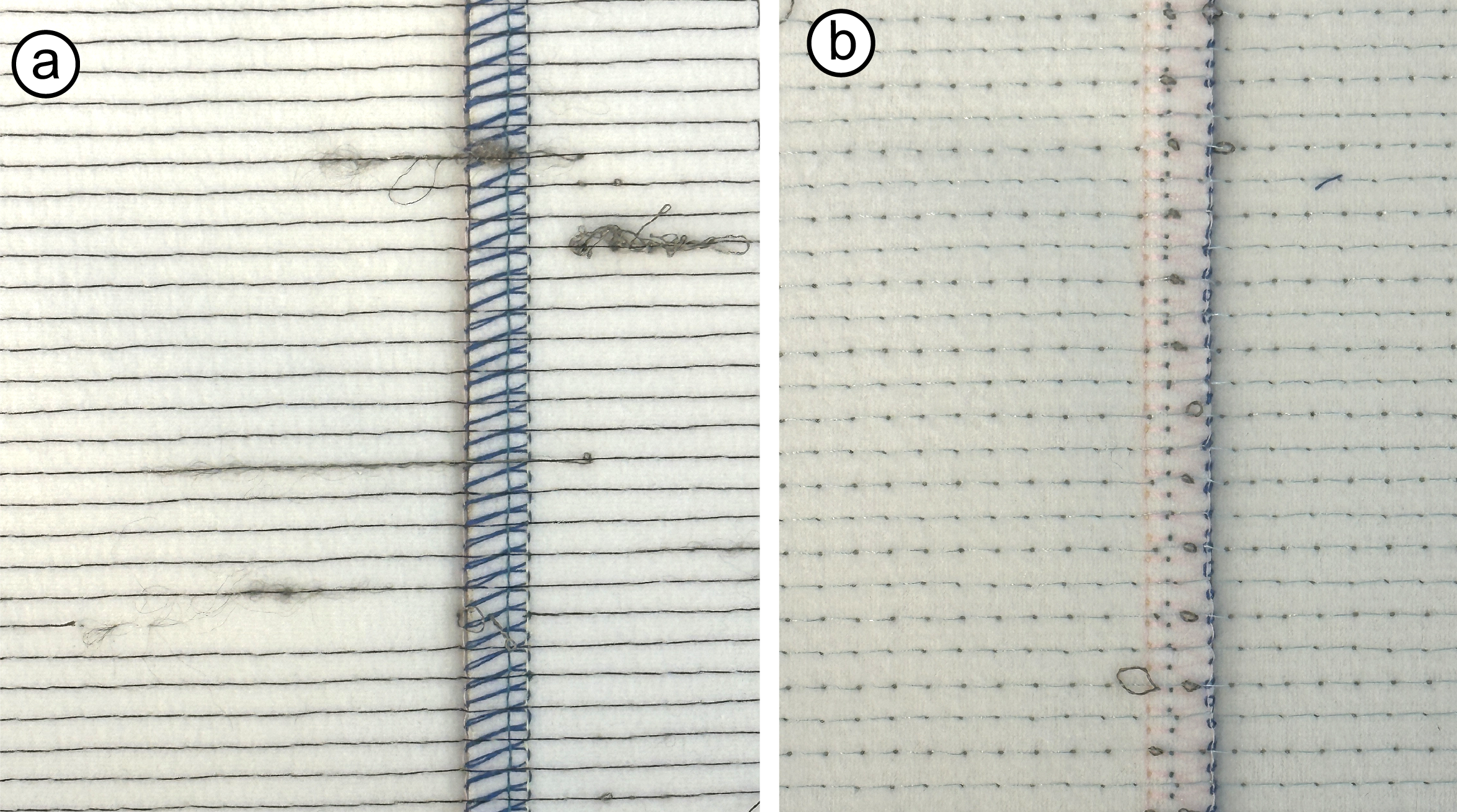}
  \caption{(a) Defects under over-tension conditions; (b) defects under well-calibrated conditions.}
  \label{fig:eval-tension-sample}
\end{figure}

The adjacent conclusion is the distinct difference in type of defects we see across variation in tension.  The high resistance variance measured in the over-tensioned condition, suggests that stitches may be electrically compromised without being visibly broken. \stitches{} reduces this variability, producing consistent resistance across tension conditions. Further inspection reveals what we see in Figure~\ref{fig:eval-tension-sample}, under over-tensioned conditions, defects appear across the \textit{entire pattern} and are not limited to seam regions. This behavior can be attributed to the uneven surface introduced by seams, which further destabilizes the already high tension, leading to inconsistent stitching conditions throughout the process. Under well-calibrated tension and under-tension, failures are localized around seam.

\subsection{Pattern orientation does not impact the results}
Since we only adjust Y-axis belt tension, we want to investigate whether this introduces directional effects. In all our experiments, the embroidery pattern is designed to cross the seam perpendicularly. We consider two pattern orientations: a horizontal pattern crossing a vertical seam, and a vertical pattern crossing a horizontal seam. While both configurations maintain a perpendicular relationship between the pattern and the seam, they differ in their alignment with respect to the machine axes.

\begin{figure}[h]
  \centering
  \includegraphics[width=\linewidth]{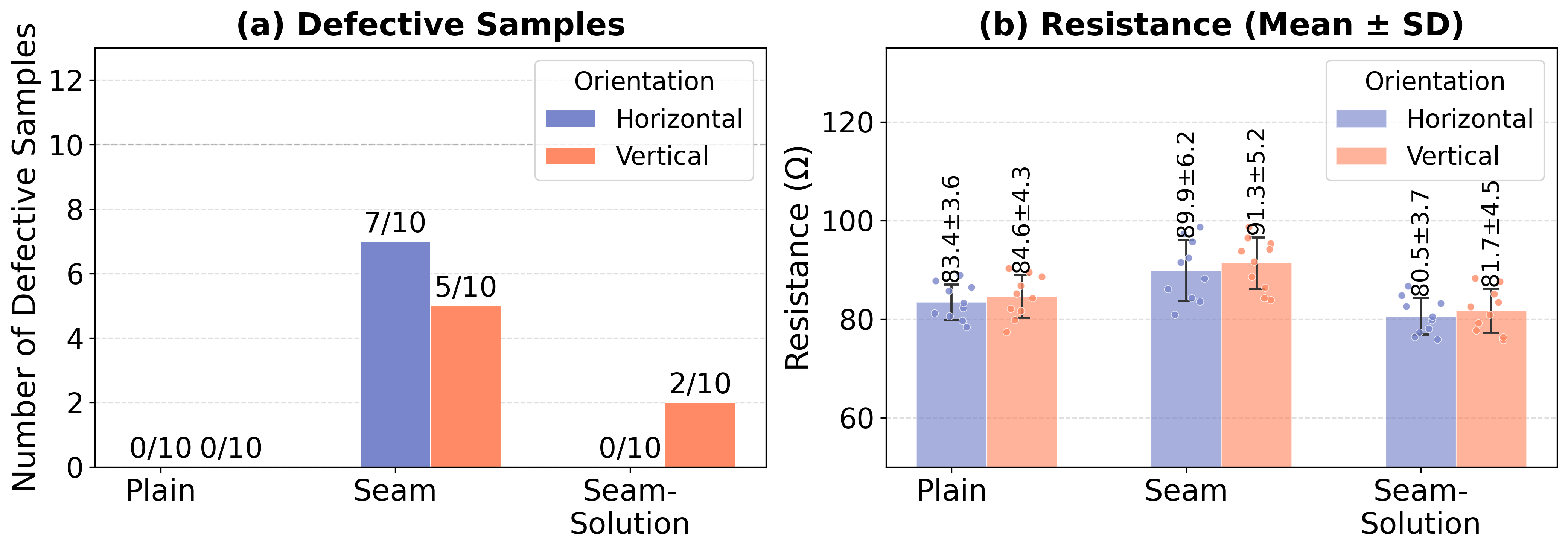}
  \caption{Effect of embroidery orientation on defect rate and trace resistance. (a)~Number of defective samples out of 10 for each construction method under horizontal and vertical orientations. (b)~Resistance measurements (mean ± SD) with individual data points.}
  \label{fig:eval-orientation}
\end{figure}

\subsubsection{Results}
As shown in Figure~\ref{fig:eval-orientation}, rotating the sensor pattern produced no meaningful change in defect rates across construction methods. \stitches{} produced 0/10 defective samples in the horizontal pattern and 2/10 in the vertical pattern. Traditional pattern digitization likewise showed similar defect counts (horizontal: 7/10, vertical: 5/10). Resistance means were also comparable across orientations for all three construction methods.

\subsubsection{Conclusion}
Pattern orientation does not significantly affect embroidery defect rates for either construction method. The remaining defects in \stitches{} samples (2/10) were limited to support penetration and floated stitches, minor defects associated with tension miscalibration rather than orientation.

\subsection{Complex seams require slight modifications}
\begin{figure}[b]
  \centering
  \includegraphics[width=\linewidth]{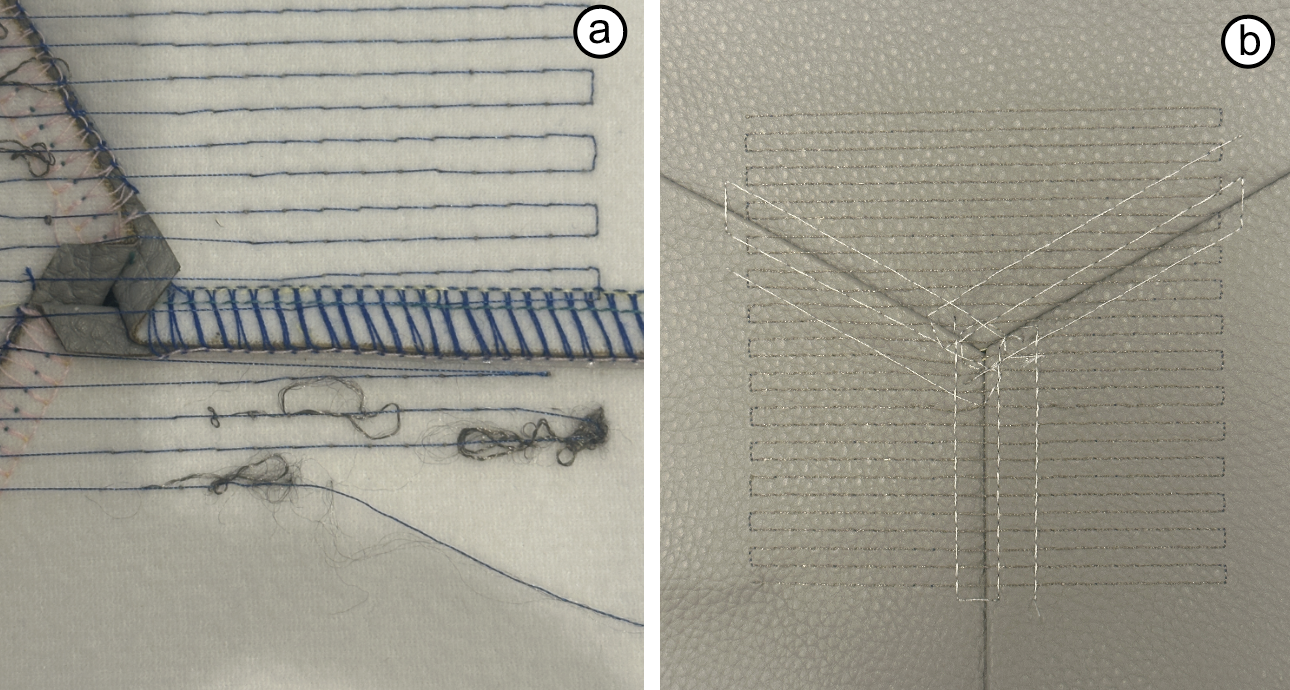}
  \caption{Three-way seam embroidering. (a)~\stitches{} solution. (b)~Global rotation solution.}
  \label{fig:three-point-seams-photo}
\end{figure}

Real world fabrics tend to come with more than a single seam, for example on sweaters/t-shirts you commonly see three-point seams on the shoulders. The different alignment of each seam with respect to the embroidery poses increased points of failure. As shown in Figure~\ref{fig:three-point-seams-photo}a, in a first naive attempt, we found that \stitches{} did not fully solve this problem because of the joint intersections. Our software tool therefore accommodates for this by giving users the choice of two solutions: (b)~a global rotation of the pattern or (c)~a slight perturbation of the needle points. Depending on the functionality of the circuit, different solutions may be desirable. We here evaluate the rotation solution.

\begin{figure}[h]
  \centering
  \includegraphics[width=\linewidth]{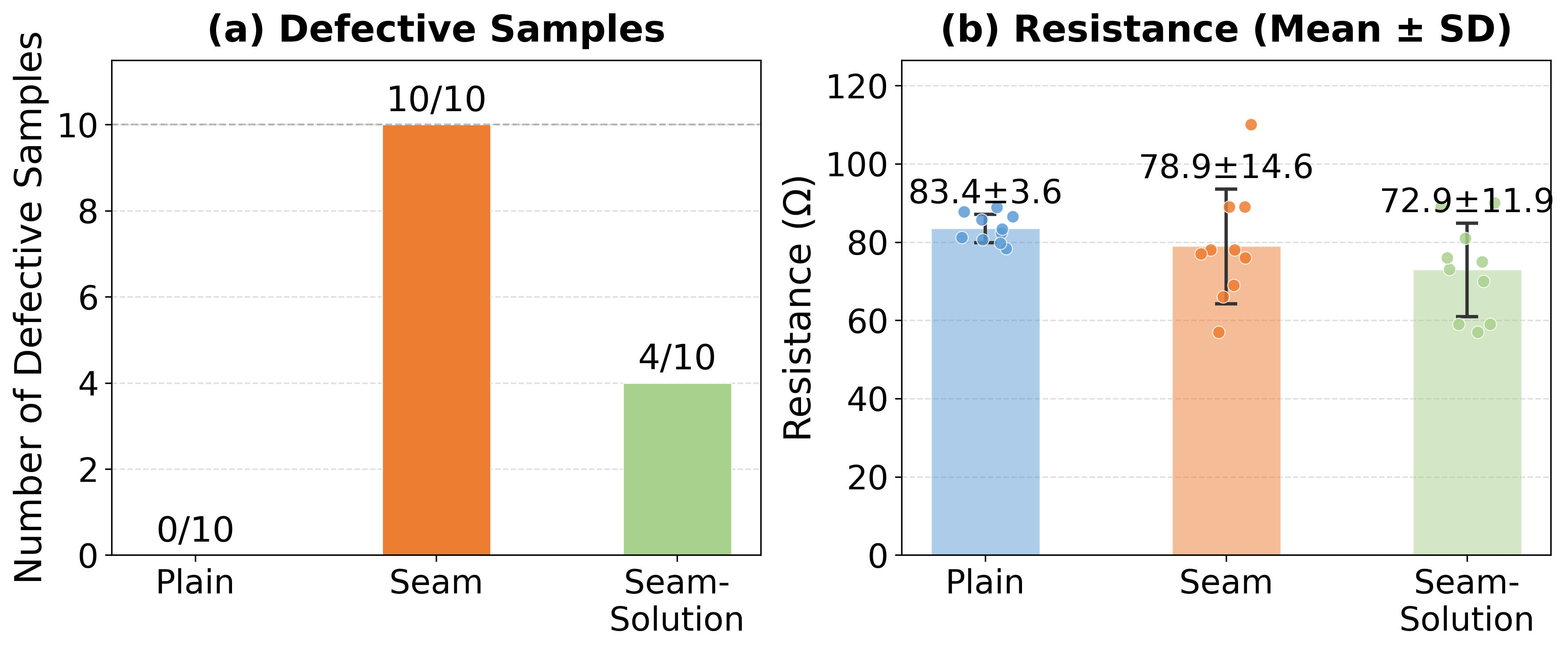
  }
  \caption{Defect rate and trace resistance for three-way seam embroidering. (a)~Number of defective samples out of 10 for each construction method. (b)~Resistance measurements (mean ± SD) with individual data points. Plain uses the well-calibrated horizontal baseline as reference.}
  \label{fig:three-point-seams}
\end{figure}

\subsubsection{Results}
Figure~\ref{fig:three-point-seams} shows the defect rate and resistance for the three-way seam condition. Without \name{}, all 10 samples exhibited defects, with thread fraying (10 samples), thread breakage (10 samples), and bird nesting (7 samples) as the dominant failure modes. \stitches{} reduced the defect rate to 4 out of 10 samples, with substantially fewer affected samples overall (thread fraying: 3, thread breakage: 2, bird nesting: 2, looping stitches: 1). Resistance in the without-\name{} condition averaged $78.9 \pm 14.6\,\Omega$ with a wide spread from $57.0$ to $110.0\,\Omega$, while \stitches{} yielded $72.9 \pm 11.9\,\Omega$ ($57.0$--$90.0\,\Omega$). Both conditions showed higher variability than the single-seam baseline (Plain: $83.4 \pm 3.6\,\Omega$).

\subsubsection{Conclusion}
The three-way seam represents a stress test for seam-crossing embroidery, as the thread must negotiate multiple intersecting fabric layers at different angles. Without \name{}, the approach fails completely (10/10 defective), producing severe defects including thread breakage and bird nesting that compromise both mechanical integrity and electrical conductivity. \stitches{} reduces the defect rate from 100\% to 40\% and substantially lowers defect frequency across all categories. 

\section{Discussion}

\subsection{Effectiveness of \stitches{}}
Through this systematic evaluation, we conclude that \stitches{} significantly reduce and often fully eliminate defects when using technical embroidery across common simple seams. As the complexity of the seam and/or the pattern increases, there is a risk that crucial needle points end up on the seams, this informed our software's suggestions for modifications either through global rotation or perturbation of the needle points. Given the potential difference in impact on functionality, we leave it up to users to decide which of these modifications they are willing to deploy. 

We furthermore find that there are distinct differences in the \textit{type of defects} across conditions, as shown in Figure~\ref{fig:defect-types-by-tension}, while not core to \stitches{} or our contribution, this finding may be insightful to diagnose early-on whether a machine is gradually getting de-calibrated, or poorly tuned to different materials, allowing for effective debugging by visually inspecting the results.

\rev{\subsection{Embroidery on non-planar textile structures}
There are three categories of topology that influence how a seamed region can be embroidered: planar seams, non-planar single-layer regions, and tubular structures. Many garment seams (e.g., side seams along the torso on shirts) can often be laid flat, allowing our method to be applied using standard embroidery practice. A non-planar region can be divided into smaller patches, each of which can be positioned as a nearly flat region and embroidered separately through multi-hooping. Floating hooping can be used for non-planar garment regions that cannot be directly hooped. For tubular structures (e.g., cuffs), the existing practice of free-arm embroidery can be used to embroider these regions. The feasibility depends on whether the tubular diameter and machine setup allow the target region to be accessed as a single-layer embroidery area. For machines without free-arm capabilities, some tubular structures can still be positioned as a single layer with floating hooping, leveraging the fabric's flexibility. }

\rev{ \subsection{Material considerations}
Our experiments focus on thick leather. Although we did not conduct a systematic materials study, exploratory tests with other fabrics suggest that seams can amplify material-specific challenges: Even on plain fabric, leather is more prone to thread breakage, while woven fabrics are prone to pattern distortion, and these material-specific issues become more pronounced on seamed fabric. More broadly, textile goods often contain material heterogeneity in composition, stretch, layering, and material combinations, all of which can affect embroidery parameters. While our tool supports region-specific settings, we did not systematically characterize parameters across diverse fabric combinations.}

\section{Application Opportunities on Textile Goods}
Drawing on prior smart textile systems, we highlight application opportunities that seam-aware technical embroidery could better support on assembled textile goods. 
Instead of confining functional embroidery to flat swatches before assembly, these examples suggest how prior applications could be placed more directly at garment locations that matter in use.

\textbf{Top-of-shoulder capacitive electrodes} could sense head-to-shoulder tapping to enable hands-free interaction, but neckline, yoke, and sleeve seams make this placement difficult to embroider directly on finished garments~\cite{geissler2024head}.
\textbf{Cuff-based capacitive touch panel} could provide direct input on garments, though folds, hems, and sleeve seams make cuffs difficult sites for technical embroidery~\cite{pourjafarian2019multi}.
\textbf{Resistive or strain-sensitive structures around joints} such as the knee, elbow, or shoulder could support stretch sensing for sports, rehabilitation, and motion monitoring, but these regions are often assembled from multiple shaped panels~\cite{martinez2021alternative}.
\textbf{Embroidered RFID tags or antennas} could support wireless functionality on garments and accessories, though their continuous routed geometry becomes difficult to fabricate when seams interrupt the path~\cite{gordon2017embroidered}.
\textbf{Sweating-sensing structures} at uniquely constructed sites, such as the underarm could also support future sweat sensing (with specialized thread), where seams and local curvature make direct embroidery especially challenging~\cite{jia2018conductive}.

\begin{figure}[h]
  \centering
  \includegraphics[width=\linewidth]{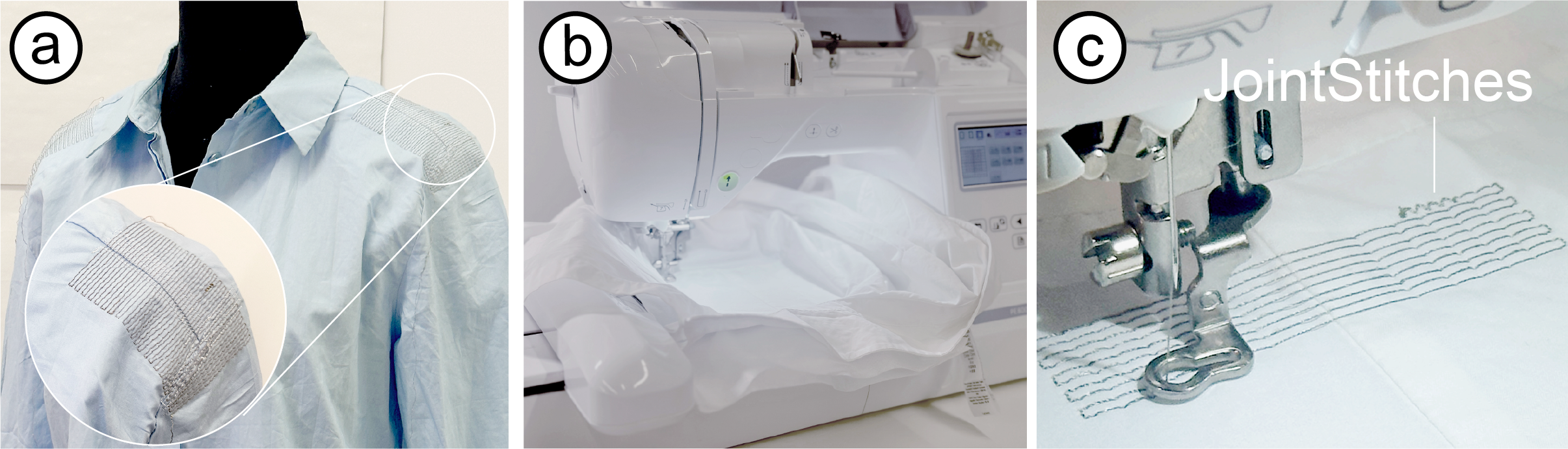}
  \caption{\rev{Top-of-shoulder capacitive electrodes embroidered on a cotton shirt for head-to-shoulder tapping detection. (a) Fabricated electrodes positioned across the Y-shaped shoulder seam. (b) Embroidery of one patch using the floating hooping technique. (c) Close-up of the automatically generated \jointstitches{} that electrically connect the conductive traces across the two separately embroidered patches.}}
  \label{fig:application}
\end{figure}

\rev{We implemented one of the proposed applications on a cotton shirt: top-of-shoulder capacitive electrodes on a Y-shaped shoulder seam for head-to-shoulder tapping detection. We used an Adafruit Circuit Playground Bluefruit together with a Texas Instruments FDC2214 to measure capacitance changes from two embroidered electrodes, one on each shoulder. For this non-planar Y-shaped shoulder seam, we divided the embroidery pattern into two patches so that each patch could be hooped as a locally flat region during embroidery. Using the floating hooping technique, the stabilizer was hooped while the shoulder area of the shirt was placed on top of the hooped stabilizer and secured without direct clamping. A pair of \jointstitches{} was automatically generated to electrically connect the separated conductive traces on the two patches.}

\section{Conclusion}
We presented the problem of seams in fabrics when it comes to technical embroidery, and presented a \stitch{} mechanism to create technical embroidery robust to seams and subtle variations in belt tension. We developed a simple software tool \name{} to support the process of inserting these \stitches{} into existing embroidery patterns. We conducted an elaborate technical evaluation to characterize the performance of existing embroidery across seams and our proposed solution, demonstrating a vast decrease (and mostly elimination) of fabrication defects caused by seams. 

While technical embroidery is a flourishing field in academic papers, it has not been frequently applied to real world fabrics and garments. We believe this is because the heterogeneous nature of these materials which deviates from the perfect patches of material on which technical embroidery typically is demonstrated. With this work we take a step towards bringing the advantages of technical embroidery to a much broader set of use cases as demonstrated in our application section, by distributing our software as open source package, we encourage the community to continue to build on this and extend what we can do with technical embroidery. The additive nature of this embroidery makes it uniquely suitable to integrate advanced functionality in a huge range of everyday fabrics. 


\bibliographystyle{ACM-Reference-Format}
\bibliography{thijs-papers,zekun-papers}

\appendix









\end{document}